\documentclass[12pt,english,nofootinbib,preprint]{revtex4-1}
\usepackage{hyperref}
\usepackage{setspace}
\usepackage{mathptmx}
\usepackage{graphicx}

\usepackage[T1]{fontenc}
\usepackage[latin9]{inputenc}
\usepackage[letterpaper]{geometry}
\usepackage{booktabs}
\usepackage{amsmath}
\usepackage{amssymb}
\usepackage{cancel}
\usepackage{babel}

\begin{document}

\title{Ground State Stability and the Nature of the Spin Glass Phase}
\begin{abstract}
We propose an approach toward understanding the spin glass phase at zero and low temperature by studying the stability of a spin glass ground state
against perturbations of a single coupling.  After reviewing the concepts of flexibility, critical droplet, and related quantities for both finite- and infinite-volume 
ground states, we study some of their properties and review three models in which these quantities are partially or fully understood. We also review a recent 
result showing the connection between our approach and that of disorder chaos. We then view four proposed scenarios for the low-temperature spin glass 
phase --- replica symmetry breaking, scaling-droplet, TNT and chaotic pairs --- through the lens of the predictions of each scenario 
for the lowest energy large-lengthscale excitations above the ground state. 
Using a new concept called $\sigma$-criticality, which quantifies the sensitivity of ground states to single-bond coupling variations, we show that each of these 
four pictures can be identified with different critical droplet geometries and energies. We also investigate necessary and sufficient conditions for the existence
of multiple incongruent ground states.
\end{abstract}
\author{C.M.~Newman}
\affiliation{ Courant Institute of Mathematical Sciences, New York University, New York, NY 10012 USA }
\affiliation{NYU-ECNU Institute of Mathematical Sciences at NYU Shanghai, 3663 Zhongshan Road North, Shanghai, 200062, China}
\author{D.L.~Stein}
\affiliation{Department of Physics and Courant Institute of Mathematical Sciences, New York University, New York, NY 10012 USA}
\affiliation{NYU-ECNU Institutes of Physics and Mathematical Sciences at NYU Shanghai, 3663 Zhongshan Road North, Shanghai, 200062, China}
\affiliation{Santa Fe Institute, 1399 Hyde Park Rd., Santa Fe, NM USA 87501}

\maketitle

\section{Introduction}
\label{sec:intro}

Despite years of intensive investigation the nature of the low-temperature phase of short-range spin glasses remains unsettled.
Among other open questions, two of the most important center on properties at zero temperature: specifically, the number of ground states and the nature of low-lying 
excitations above them.  Ground state multiplicity is an essential feature required for the correct characterization of the zero-temperature thermodynamics of spin glasses 
(and perhaps counterintuitively, plays an important role in influencing nonequilibrium dynamics following a deep quench~\cite{NS99a,NS99c,NS08,JRY21}), while
large-lengthscale, low-energy excitations above the ground state at zero temperature are important in determining the thermal properties of the spin glass phase
at low but nonzero temperatures.

In this paper we introduce and develop a new approach to understanding ground state properties, by focusing on ground state stability with respect to coupling perturbations. 
Some of the concepts used in this paper have appeared in earlier work~\cite{NS2D00,NS2D01,ADNS10,ANS19}, but so far they have not been integrated into a unified picture. 

We will see that the concept of disorder chaos~\cite{BM87,FH88b,KB05,KK07,Chatterjee09,ANS19} is related to the approach used here, though disorder chaos focuses on the behavior of a ground state when {\it all\/} of the coupling values have changed by a small, usually random, amount.  In this paper we focus instead on the behavior of a ground state when a {\it single\/} 
coupling value is changed --- in a direction to destabilize the ground state --- by an~arbitrary amount. The first question that naturally arises is, by how much does one need to change the coupling value at an arbitrarily chosen edge to cause a droplet flip (i.e., the uniform reversal of a subset of spins in an Ising model)? If the coupling distribution has finite, $O(1)$-variance one can easily prove an $O(1)$ upper bound for the change required.  

So if we then destabilize the ground state by making an $O(1)$ change to a single coupling in an infinite system, how large can the resulting change be? One might naturally expect that any such change should be local, involving the overturning of $O(1)$ spins. This is almost certainly true for most edges (if not all) in the Edwards-Anderson (EA) spin glass~\cite{EA75} in any dimension greater than one.  (We will see however in Sect.~\ref{subsec:special} that this is not true in one dimension for {\it any\/} edge, but for a trivial reason special to $1D$.)

But a droplet flip containing~$O(1)$ spins need not necessarily be the case for all edges in all dimensions. In fact, the most interesting case to consider is one where an $O(1)$ change in the coupling value at a single edge leads to a {\it complete transformation\/} of an infinite-volume ground state~$\sigma$, to a new ground state~$\sigma'$ with nontrivial spin {\it and\/} edge overlap with $\sigma$. This property is one example of what we will call ``$\sigma$-criticality''.

The question then becomes, could such edges even exist? We will see not only that as of now the answer is yes in principle, but that they {\it must\/} occur in any dimension in which the replica symmetry breaking~(RSB) picture might hold. 

As already noted, the method used in this paper is to consider a spin glass ground state when all couplings are held fixed but one. As we've discussed elsewhere~\cite{NS2D00,NS2D01,ADNS10,ANS19} (but will review in Section~\ref{sec:droplets}), in any ground state every edge has a single coupling value, called its {\it critical value\/}, which is determined by all of the other coupling values {\it except\/} for that associated with the edge in question. On either side of an edge's critical value there are two distinct ground states differing by a droplet flip whose boundary passes through that edge.  We refer to this as the {\it critical droplet\/} of the edge in that particular ground state.

For a fixed coupling realization, and a specific ground state consistent with that realization, every coupling value is a certain distance (on the real line) from its edge's critical value; we refer to this 
distance as the coupling's {\it flexibility\/}~\cite{NS2D00,NS2D01,ADNS10,ANS19}.  A ground state can be characterized by the collection of all its flexibilities; these and critical droplet size distributions will be the objects of our study. 

\bigskip

The plan of this paper is as follows: in Sect.~\ref{sec:droplets} we introduce basic concepts and definitions related to finite-volume ground states, critical droplets, and flexibilities and discuss some of their properties. In~Sect.~\ref{sec:inf} we introduce the zero-temperature periodic boundary condition metastate and extend the discussion of Sect.~\ref{sec:droplets} to infinite systems. We also discuss three spin glass models in which critical droplet geometries are completely or partially understood. In Sect.~\ref{sec:chaos} we examine the connection between critical droplets, disorder chaos, energy fluctuations, and ground state multiplicity. In Sect.~\ref{sec:interfaces} we review some of the main conjectured scenarios for the low-temperature spin glass phase and classify them in terms of their predicted lowest-energy large-lengthscale excitations.

Sections~\ref{sec:droplets} through~\ref{sec:interfaces} provide the necessary background, concepts, and terminology needed for subsequent sections. Although some of the material in Sects.~\ref{sec:droplets}--\ref{sec:interfaces} is new, most of the results cited in those sections have been published, though in a scattered, piecemeal fashion. The remaining sections consist entirely of new results. In Sect.~\ref{sec:sfcd} the concept of $\sigma$-criticality is defined and it is proved that this property holds for space-filling critical droplets. In Sect.~\ref{sec:consequences} we examine some consequences of the possible presence of $\sigma$-criticality in spin glass ground states, including its effect on ground state multiplicity. In addition, a new translation-invariant measure for classifying ground states is proposed and its properties studied. In Sect.~\ref{sec:rsb} we establish the connection between $\sigma$-criticality and interfaces between ground states in the replica symmetry breaking picture of the spin glass phase, and use this to prove a theorem on ground state multiplicity in the RSB picture. In Sect.~\ref{sec:multi} we use $\sigma$-criticality to find necessary and sufficient conditions for ground state multiplicity to arise more generally. We summarize our results in Sect.~{\ref{sec:discussion} with a focus on how each of the four pictures of the spin glass phase presented in Sect.~\ref{sec:interfaces} corresponds to its own category of critical droplet geometries and, through the concept of $\sigma$-criticality, type of ground state stability/instability.


\section{Ground states and critical droplets}
\label{sec:droplets}

In this paper we focus on the Edwards-Anderson (EA) nearest-neighbor Ising spin glass model~\cite{EA75} in zero magnetic field on the $d$-dimensional cubic lattice $\mathbb{Z}^d$:
\begin{equation}
\label{eq:EA}
{\cal H}_J=-\sum_{<x,y>} J_{xy} \sigma_x\sigma_y 
\end{equation}
where $\sigma_x=\pm 1$ is the Ising spin at site $x$ and $\langle x,y\rangle$ denotes an edge (or ``bond'' --- we will use the two terms interchangeably) in the (nearest-neighbor) edge set $\mathbb{E}^d$. The couplings $J_{xy}$ are independent, identically distributed continuous random variables chosen from a distribution $\nu(dJ_{xy})$, with random variable $J_{xy}$  assigned to the edge $\langle x,y\rangle$.  Our requirements on $\nu$ is that it be supported on the entire real line, is distributed symmetrically about zero, and has finite variance; e.g., a Gaussian with mean zero and variance one.  We denote by $J$ a particular realization of the couplings.

A finite-volume ground state is the lowest-energy spin configuration in a finite volume~$\Lambda_L$ with a specified boundary condition; here the sum in~(\ref{eq:EA}) is restricted to edges touching at least one site entirely within $\Lambda_L$ (that is, the sum includes not only edges with both sites entirely within $\Lambda_L$ but also edges having one site in $\Lambda_L$ and a nearest-neighbor site on the boundary $\partial\Lambda_L$). We will always choose $\Lambda_L$ to be a cube of side~$L$ centered at the origin. If the boundary condition is spin-symmetric, such as free or periodic, then given the spin-flip symmetry of the EA~Hamiltonian, ground states will come in spin-reversed pairs; we discuss this further in Sect.~\ref{subsec:pbc}.  

An {\it infinite-volume\/} ground state is defined by the condition that its energy cannot be lowered by flipping any {\it finite\/} subset of spins. That is, if $\sigma$ denotes an infinite-volume ground state (always defined with respect to a specific~$J$, though this dependence will be suppressed for notational convenience), then 
\begin{equation}
\label{eq:gs}
E_{S}=\sum_{<x,y>\in{S}}J_{xy}\sigma_x\sigma_y\ >0 
\end{equation}
where $S$ is any closed $(d-1)$-dimensional surface (or contour in two dimensions) in the dual lattice; i.e., $S$ is a surface completely enclosing a connected set of spins (a ``droplet''). 

(One quick remark about notation before moving on: strictly speaking, it is the dual bond $\langle x,y\rangle^*$, not $\langle x,y\rangle$ itself, that belongs to $S$, which is a surface in the dual lattice. However, because the sum in~(\ref{eq:gs}) is over sites and edges in the original lattice, we abuse notation somewhat and write $\langle x,y\rangle\in S$ in the sum. This should be understood as meaning, ``sum over edges in the original lattice whose duals belong to $S$.'')

The inequality in~(\ref{eq:gs}) is strict since, by the continuity of $\nu(dJ_{xy})$, for a pair $({J},\sigma)$ there is zero probability of any closed surface having exactly zero energy in $\sigma$. The condition~(\ref{eq:gs}) must also hold for finite-volume ground states for any closed surface completely inside $\Lambda_L$. It is then not hard to show that an alternative (and equivalent) definition, which we also sometimes use, is that an infinite-volume ground state is any convergent limit of an infinite sequence of finite-volume ground states.

The notions of critical droplets and flexibilities were introduced in~\cite{NS2D00} and arise most naturally from the construction of the {\it excitation metastate\/}, also introduced in~\cite{NS2D00} (see also~\cite{NS2D01,ADNS10,ANS19}). Here we avoid technicalities and introduce these concepts without bringing in the excitation metastate; the reader is referred to the above references for a fuller discussion. (For a complete and fully rigorous definition of critical droplets and related notions using the excitation metastate, see~\cite{NS2D00,NS2D01,ADNS10,ANS19}).  For fixed coupling realization $J$, consider a finite-volume ground state $\sigma^>$ and a specific edge $\langle x,y\rangle$ with coupling value $J_{xy}$. For ease of discussion suppose $J_{xy}=K>0$ in $J$ and that it is satisfied in $\sigma^>$.  Hold all coupling values fixed except for $J_{xy}$. For increasing values of $J_{xy}>K$, $\sigma^>$ only becomes more stable and remains unchanged. The ground state $\sigma^>$ will also remain unchanged (though with decreasing stability) for a finite range of values of $J_{xy}$ {\it below\/} $K$. Eventually, below some (positive or negative) value $J_{xy}=J_c<K$, the ground state becomes unstable and a droplet of spins will overturn, leading to a new ground state $\sigma^<$. It is easy to see that decreasing $J_{xy}$ further below $J_c$ now {\it increases\/} the stability of $\sigma^<$. 

The conclusion is that this procedure --- i.e., all couplings but $J_{xy}$ held fixed and $J_{xy}$ varying from $-\infty$ to $+\infty$ --- leads to a pair of ground states $\sigma^>$ and $\sigma^<$, differing by a droplet flip, as follows:  there is a critical value $J_c$, determined by all couplings {\it except\/} $J_{xy}$, such that for $J_{xy}>J_c$, the ground state is $\sigma^>$, while for $J_{xy}<J_c$, the ground state is $\sigma^<$.

What happens exactly at $J_c$? It is not hard to see that precisely at that value, in both $\sigma^>$ and $\sigma^<$, there will be a (shared) unique closed surface ${S}_0$ in the dual lattice which includes the dual edge $\langle x,y\rangle^*$ and has precisely zero energy as defined by~(\ref{eq:gs}), with every other surface in the dual lattice having strictly positive energy.  

We refer to the volume (or equivalently, set of spins), enclosed by this zero-energy (precisely at $J_c$) surface ${S}_0$ as the {\it critical droplet\/} of $\langle x,y\rangle$ in $({J},\sigma^>)$. For our original~$({J},\sigma^>)$, the critical droplet boundary ${S}_0$ is the closed surface which includes $\langle x,y\rangle^*$ with least energy $E_{{S}_0}$ in the original $J$. It follows that, in {\it both\/} $\sigma^>$ and $\sigma^<$,  ${S}_0$ is the lowest-energy dual lattice surface containing $\langle x,y\rangle^*$ for {\it any\/} value of $J_{xy}$ (as always, with all other couplings held fixed to their original values in $J$). Moreover, though $E_{{S}_0}$ depends on the value $J_{xy}$, it is always positive in the corresponding ground state (i.e., $\sigma^>$ or $\sigma^<$), except at $J_c$ where it is zero in both. 

While the discussion above assumed that $J_{xy}$ was originally positive and satisfied in $\sigma$, the conclusions hold in general.
A more formal treatment~\cite{NS2D00,NS2D01} using metastates shows that all of the properties described above survive in the infinite-volume limit and therefore hold also for infinite-volume ground states; the only difference (but an important one!) is that in infinite-volume ground states a critical droplet can be either finite or infinite in extent. Any finite critical droplet consists of a connected set of spins, but that need not be true in the infinite volume limit. We defer further discussion of infinite critical droplets (which play an important role in what follows) to Sect.~\ref{subsec:cdinf}.

\bigskip

Based on this discussion, we can now define the concepts of critical droplets and flexibilities. We begin by considering the EA model in a finite volume $\Lambda_L$.

\medskip

{\bf Definition 2.1.} Consider the ground state (or ground state pair, to be discussed in Sect.~\ref{subsec:pbc}) $\sigma_L$ for the EA Hamiltonian~(\ref{eq:EA}) on a finite volume $\Lambda_L$ with boundary conditions chosen independently of $J$. (For ease of discussion we restrict attention to simple, commonly used boundary conditions, such as free, periodic, antiperiodic, or fixed). Choose an edge (or bond) $b_{xy}=\langle{x,y}\rangle$ with $x,y\in\Lambda_L$.  Consider all closed surfaces in the dual edge lattice ${\mathbb E}_L^*$ which include the dual bond $b^*_{xy}$. By~(\ref{eq:gs}) these all have positive energy; moreover, because of the continuity of $\nu(dJ_{xy})$ there are no ``ties'' in these energies. Therefore there exists a closed surface $\partial D(b_{xy},\sigma_L)$, passing through $b^*_{xy}$, of {\it least\/} energy in $\sigma_L$. We call $\partial D(b_{xy},\sigma_L)$ the {\it critical droplet boundary\/} of $b_{xy}$ in $\sigma_L$ and the set of spins $D(b_{xy},\sigma_L)$ enclosed by $\partial D(b_{xy},\sigma_L)$ the {\it critical droplet\/} of $b_{xy}$ in $\sigma_L$.

\medskip

{\bf Remark.} Critical droplets are defined with respect to edges rather than associated couplings to avoid confusion, given that we will often vary the coupling value associated with specific edges, while the edges or bonds themselves are fixed, geometric objects.

\medskip

{\bf Definition 2.2.} The {\it energy\/} $E\bigl(J,D(b_{xy},\sigma_L)\bigr)$ of the critical droplet of $b_{xy}$ in $\sigma_L$ for coupling realization $J$ is defined to be the energy of its boundary as given by~(\ref{eq:gs}):
\begin{equation}
\label{eq:cd}
E\bigl(J,D(b_{xy},\sigma_L)\bigr)=\sum_{<x,y>\in\partial D(b_{xy},\sigma_L)}J_{xy}\sigma_x\sigma_y\, .
\end{equation}

\medskip

{\bf Definition 2.3.} The {\it critical value\/} of the coupling $J_{xy}$ associated with $b_{xy}$ in $\sigma_L$ is defined as the value of $J_{xy}$ where $E\bigl(J,D(b_{xy},\sigma_L)\bigr)=0$, while all other couplings in $J$ are held fixed.

\medskip

{\bf Remark on notation.} As mentioned earlier, the critical value of the coupling $b_{xy}$ in $\sigma_L$ is determined by all of the couplings in ${\mathbb E}_L$ {\it except\/} $J_{xy}$; for further discussion on this point, see~\cite{NS2D00,NS2D01,ADNS10,ANS19}. In later sections we will use this fact to study how the critical droplet energy of $b_{xy}$ changes when $J_{xy}$ is varied while keeping all other couplings fixed. For this reason, to avoid confusion we hereafter drop the explicit dependence of droplet energy on~$J$, which is understood, and simply write~$E\bigl(D(b_{xy},\sigma_L)\bigr)$.

\smallskip

{\bf Remark.} The definition of critical droplets is not restricted to closed surfaces entirely within $\Lambda_L$; i.e., it is possible for a critical droplet to reach the boundary $\partial\Lambda_L$, with the proviso that the droplet, if overturned, must still obey the imposed boundary conditions. Hence a critical droplet reaching the boundary is ruled out for fixed boundary conditions but is allowed for free, periodic, or antiperiodic boundary conditions. In the case of free boundary conditions, a critical droplet reaching the boundary will not be a closed surface within $\Lambda_L$ (excluding $\partial\Lambda_L$); if it touches two separate faces of $\partial\Lambda_L$ it would then divide the spins in $\Lambda_L$ into two disjoint components both of which extend to the boundary. For periodic boundary conditions, the critical droplet boundary is a closed surface enclosing a connected droplet of spins in the equivalent $d$-dimensional torus, but both surface and droplet can appear disconnected when viewed within the cube $\Lambda_L$.

We complete this section by introducing the flexibility $f(J_{xy},\sigma_L)$ for fixed coupling realization $J$.

\medskip 

{\bf Definition 2.4.} For fixed $J$, let $J_{xy}$ be the coupling value of the bond~$b_{xy}$ and $J_c(b_{xy},\sigma_L)$ be the critical value of $b_{xy}$ in $\sigma_L$.  We define the flexibility $f(b_{xy},\sigma_L)$ of $b_{xy}$ in $\sigma_L$ to be $f(b_{xy},\sigma_L)=|J_{xy}-J_c(b_{xy})|$. (Dependence of flexibility on $J$ is suppressed for ease of notation, but is understood.)

\medskip

{\bf Remark.} The flexibility, as briefly discussed in the Introduction, is the distance of a coupling's value in a realization $J$ from its critical value in a specified ground state. It is therefore in some sense a measure of the stability of $\sigma_L$. More precisely, it measures the sensitivity of a ground state at a particular edge as a function of the couplings: the larger the flexibility, the more stable is the ground state under changes of the edge's coupling value. Hence the flexibility, as opposed to the critical droplet, is defined with respect to couplings rather than edges because, unlike its critical droplet, the flexibility of an edge depends on its associated coupling value in $J$. (Note: the definition of flexibility given here differs by a factor of two from that given in~\cite{ANS19}, where flexibility of a coupling was defined as the energy cost of its critical droplet flip in the original $J$.)

\smallskip

{\bf Remark.} It was noted following the definition of the critical value $J_c$ of a bond $b_{xy}$ with coupling value $J_{xy}$ that $J_c$ is determined by all couplings in $J$ {\it except\/} $J_{xy}$.  Because couplings are chosen independently from $\nu(dJ_{xy})$, it follows that the value $J_{xy}$ is {\it independent\/} of~$J_c$. Therefore, given the continuity of $\nu(dJ_{xy})$, there is zero probability in a ground state that any coupling has exactly zero flexibility.

\smallskip

It follows from the definitions above that
\begin{equation}
\label{eq:flex}
f(b_{xy},\sigma_L)=E\bigl(D(b_{xy},\sigma_L)\bigr)\, .
\end{equation}
Therefore couplings which share the same critical droplet have the same (strictly positive) flexibility. 

From Definitions~2.1 -- 2.4 some simple properties of critical droplets and flexibilities can be immediately deduced. We collect these in the following two lemmas, which will be useful in later sections.

\medskip

{\bf Lemma 2.5.}  Consider two distinct edges $b_1$ and $b_2$ and a finite-volume ground state $\sigma_L$. Then:

\smallskip 

(a) If $f(b_1,\sigma_L) > f(b_2,\sigma_L)$, then $b_1$ cannot belong to $\partial D(b_2,\sigma_L)$. ($b_2$ may or may not belong to the critical droplet boundary of~$b_1$.)

\smallskip

(b) If $b_1\in\partial D(b_2,\sigma_L)$ and $b_2\in\partial D(b_1,\sigma_L)$, then $b_1$ and $b_2$ share the same critical droplet.  Equivalently, if two bonds have equal flexibilities, they share the same critical droplet.

\medskip

{\bf Proof.} (a) This is an elementary consequence of~(\ref{eq:flex}) combined with the fact that for any edge~$\langle x,y\rangle$, its critical droplet boundary is the closed surface of minimum energy that includes $\langle x,y\rangle$.

\smallskip

(b) From (a), $b_1\in\partial D(b_2,\sigma_L)$ implies that  $f(b_1,\sigma_L) \le f(b_2,\sigma_L)$, and $b_2\in\partial D(b_1,\sigma_L)$ implies that  $f(b_2,\sigma_L) \le f(b_1,\sigma_L)$. Therefore $f(b_1,\sigma_L) = f(b_2,\sigma_L)$. Using~(\ref{eq:flex}), we then have $E\bigl(D(b_1,\sigma_L)\bigr)=E\bigl(D(b_2,\sigma_L)\bigr)$.  But by the continuity of the coupling distribution, there is zero probability that any two distinct bounded surfaces have identical energies. The result then follows.

\bigskip

{\bf Lemma 2.6.} If the flexibility of any edge is lowered (by changing its coupling value) but remains positive in $\sigma_L$, the flexibility of any other edge in $\sigma_L$ is either also lowered (by the same amount) or else remains unchanged.

\medskip

{\bf Proof.}  Choose an arbitrary bond $b_0$ with coupling value $J_0$ in $J$ and critical value $J_c$ in $\sigma_L$.  Without loss of generality let $J_0>J_c$. Changing the initial coupling value $J_0$ to a lower value $J(b_0)$ with $J(b_0)\in(J_c,J_0)$ lowers the flexibility of $b_0$ without affecting $\sigma_L$. The question then becomes whether it affects the flexibilities of other bonds in $\sigma_L$. Changing the value of $J(b_0)$ can affect only the flexibilities of bonds $b_i$ with $b_0\in\partial D(b_i,\sigma)$, so let us consider such a bond. Using~(\ref{eq:cd}) we have
\begin{equation}
\label{eq:lower}
E\bigl(D(b_i,\sigma_L)\bigr)=\sum_{<x,y>\in\partial D(b_i,\sigma_L)}J_{xy}\sigma_x\sigma_y=J(b_0) + \sum_{<x,y>\in\partial D(b_i,\sigma_L)\backslash b_0}J_{xy}\sigma_x\sigma_y\, .
\end{equation}
Using~(\ref{eq:flex}) the flexibility $f(b_i,\sigma_L)$ is then
\begin{equation}
\label{eq:flex2}
f(b_i,\sigma_L)=E\bigl(D(b_i,\sigma_L)\bigr)=J(b_0) + \sum_{<x,y>\in\partial D(b_i,\sigma_L)\backslash b_0}J_{xy}\sigma_x\sigma_y\, .
\end{equation}
Therefore, in the process of lowering $J(b_0)$ without passing its critical value, $f(b_i,\sigma_L)$ is also lowered without any change in $\sigma_L$, and the amount by which it is lowered is the is the same for any $b_i$ satisfying $b_0\in\partial D(b_i,\sigma)$. This proves the lemma.

\bigskip

The discussion above has been confined to finite-volume ground states. Before extending these ideas to infinite-volume ground states, it is helpful to introduce the periodic boundary condition metastate, which will be done in the following section.

\section{Metastates and critical droplets in infinite-volume ground states}
\label{sec:inf}

\subsection{The zero-temperature periodic boundary condition metastate}
\label{subsec:pbc}

We have argued in earlier papers that a natural setting for studying the equilibrium thermodynamics of inhomogeneous systems is the {\it metastate\/}~\cite{AW90,NS96c,NSBerlin,NS97,NS98,NS03b,Read14}, and we will use that setting here. The metastate of interest in this paper is the {\it zero-temperature periodic boundary condition metastate\/}, denoted $\kappa_J(\sigma)$ (or often simply $\kappa_J$), which is a probability measure on infinite-volume ground states $\sigma$ induced by an infinite sequence of volumes with periodic boundary conditions. This ``ground-state metastate'' is a simpler construct than the excitation metastate mentioned in Sect.~\ref{sec:droplets}, which contains all thermodynamic information that can be generated in the ensemble of all volumes and which as noted earlier will be omitted from the present discussion.

A quick note before proceeding:  because both periodic conditions and  the Hamiltonian~(\ref{eq:EA}) obey spin-flip symmetry, the finite-volume ground states generated in each volume appear as spin-reversed pairs $\sigma_L$ and $\overline{\sigma_L} (=-\sigma_L$), so henceforth a general ground state label $\alpha$ should be understood to refer to a mixed thermodynamic state consisting of a {\it pair\/} of spin-reversed ground states $(\alpha,\overline{\alpha})$ each with weight 1/2. It is easily seen that for any edge, any spin-reversed pair of ground states share the same critical droplet and flexibility.

There are two independent constructions of metastates, one initially constructed for random-field magnets~\cite{AW90} and one initially constructed for spin glasses~\cite{NS96c}; both constructions are sufficiently general that they can be used for a wide variety of applications, such as mean-field Curie-Weiss ferromagnets with random couplings~\cite{Kuelske97,Kuelske98}, neural networks~\cite{Kuelske97,vES01}, and other disordered systems. It was proved in~\cite{NSBerlin} that there exists an infinite sequence of volumes, chosen independently of $J$, for which the two constructions give an identical metastate, and so either method can be used to construct the $\kappa_J$ introduced above. 

In the approach of~\cite{AW90}, one considers for each $\Lambda_L$ with periodic boundary conditions the random pair $(J_L,\sigma_L)$, where $J_L$ is the restriction of $J$ to
$\mathbb{E}_L$, and takes the limit (using compactness) of these finite-dimensional distributions along a $J$-independent subsequence
of $L$'s. This yields a probability distribution $\kappa$ on infinite-volume $(J,\sigma)$'s which (given the use of PBC's) is translation-invariant under 
simultaneous lattice translations of $J$ and $\sigma$. The ground state metastate is then the conditional distribution $\kappa_J$ of $\kappa$ given a fixed $J$, 
and is supported entirely on GSP's for that $J$. 

The physical nature of the metastate is perhaps more transparent using the alternative construction of~\cite{NS96c}. Consider an infinite sequence of volumes $\Lambda_{L_1},\Lambda_{L_2}\ldots$ all centered at the origin and with with $L_1\ll L_2\ll\ldots$ such that $L_k\to\infty$ as $k\to\infty$.  The zero-temperature PBC metastate $\kappa_J$ is then defined through the following construction: given the sequence of volumes introduced above we can construct a type of microcanonical ensemble $\kappa_N$ in which each of the finite-volume ground state pairs $\sigma_{L_1},\sigma_{L_2},\ldots,\sigma_{L_N}$ has weight $N^{-1}$. The ensemble $\kappa_N$ converges to the metastate $\kappa_J$ as $N\to\infty$ in the sense that, for every well-behaved function $g$ on ground states (e.g., a function on finitely many spins),
 \begin{equation}
\label{eq:metastate}
\lim_{N\to\infty} N^{-1}\sum_{k=1}^N g(\sigma_{L_k})=\int g(\sigma)\ d\kappa_J(\sigma)\, .
\end{equation}
From~(\ref{eq:metastate}) we see that $\int d\kappa_J(\sigma)=1$ and $\kappa_J$ can therefore be interpreted as a probability measure on ground state pairs: the finite-volume probability of any event depending on a finite set of spins and/or couplings converges in the infinite-volume limit to the $\kappa_J$-probability of that event.

The information contained in $\kappa_J$ includes the fraction of cube sizes $L_k$ which the system spends in different {\it infinite-volume ground state pairs\/} $\sigma$ as $k\to\infty$. By this we mean the following: choose a fixed ``window'', i.e., a cube $\Lambda_w$ of side $w$ centered at the origin. If there is only a single pair of ground states, as in the droplet-scaling model~\cite{Mac84,BM85,FH88b,FH86}, then the spin configurations in $\Lambda_w$ generated by the finite-volume ground state pairs will eventually settle down to a fixed configuration, which is the restriction of the infinite-volume GSP to $\Lambda_w$~\cite{NS92}. On the other hand, if there are many infinite-volume ground state pairs, as in the replica-symmetry-breaking (RSB) picture~\cite{Read14,Parisi79,Parisi83,MPSTV84a,MPSTV84b,MPV87,MPRRZ00}, then the spin configuration in $\Lambda_w$ never converges to a limit (a phenomenon we have called {\it chaotic size dependence\/}~\cite{NS92}). Instead, for any $\Lambda_{L_k}$ with $L_k$ sufficiently large, the pair of spin configurations in $\Lambda_w$ will be identical to that of one of the many infinite-volume ground states pairs available for the system to choose from, and the ``chosen'' GSP varies with~$L_k$. Although the spin configuration in $\Lambda_w$ never settles down, the {\it fraction of volumes\/} $\Lambda_{L_k}$ for which the local spin configuration (i.e., in any fixed window) of any particular ground state pair appears {\it does\/} converge to a limit, and this information is contained within~$\kappa_J$. For a fully rigorous treatment, see~\cite{NSBerlin}. 

\subsection{Critical droplets in infinite-volume ground states}
\label{subsec:cdinf}

We now extend the discussion of Sect.~\ref{sec:droplets} to infinite-volume ground states. As noted earlier, a rigorous definition of critical droplets and flexibilities requires use of the excitation metastate~\cite{NS2D00,NS2D01,ADNS10,ANS19}, but we will define these and related quantities informally here and refer the interested reader to the references for a complete discussion. The main point is that finite-volume critical droplets and flexibilities converge with their properties preserved in the infinite-volume limit: sequential compactness of the finite-volume ground states leads to convergence 
along deterministic subsequences of $\Lambda_L$'s of the associated finite-dimensional probability distributions (involving finitely many couplings and spins) to a limiting translation-invariant probability measure $\kappa_J$ on infinite-volume configurations. The relative compactness for ground states and critical droplets follows from the two-valuedness of the Ising spins and that for the critical value follows from the ($L$-independent) trivial bound 
\begin{equation}
\label{eq:flexbound}
\vert J_c(b_{xy},\sigma_L)\vert\le{\rm min}(\sum_{z\ne y\atop |z-x|=1}\vert J_{xz}\vert,\sum_{u\ne x\atop |y-u|=1}\vert J_{uy}\vert)
\end{equation}
for any edge $\langle x,y\rangle$.

From this discussion we present the following lemma, omitting a formal proof.

\medskip

{\bf Lemma 3.1.}  All of the properties of critical droplets listed in Lemmas~2.5 and 2.6 are preserved in the infinite volume limit.

\medskip

The above discussion would not be needed in an informal treatment if all critical droplets in infinite-volume ground states were finite, i.e., completely bounded within a finite volume; they can then be defined exactly as in~Sect.~\ref{sec:droplets}. (In fact, it was proved~\cite{ANS21} that this is the case in two spin glass models; we will return to this in Sect.~\ref{subsec:special}.) However, the possibility that critical droplets can be infinite in extent in one or more directions must also be considered. Here metastates are a convenient tool for defining such unbounded critical droplets, comprising an infinite subset of spins: they are the infinite-volume limits of critical droplets in finite-volume ground states.

How might such unbounded critical droplets arise? One possibility (there may be others) is the following: consider a fixed edge $b_{xy}$ whose finite-volume ground state critical droplets increase in size as $L$ increases, while the corresponding flexibilities decrease as $L$ increases. In the limit $L\to\infty$ one then arrives at a critical droplet with infinite boundary comprising an infinite subset (with respect to $\mathbb{Z}^d$) of spins and with a well-defined (and still strictly positive) limiting flexibility. We will be particularly interested in a special case of unbounded critical droplets:

\medskip

{\bf Definition 3.2.} Consider an edge $b_{xy}$ and an infinite-volume ground state $\sigma$. We will say that  ``the critical droplet of $b_{xy}$ in $\sigma$ is space-filling'' to mean that $\partial D_{b_{xy},\sigma}$ comprises a positive density of bonds in $\mathbb{E}^d$.

\medskip

We will prove in Sect.~\ref{sec:rsb} that space-filling critical droplets are an essential component of the RSB picture of the low-temperature spin glass phase. At present we do not know whether space-filling critical droplets exist at all in the~EA spin glass or, if they do, whether their existence is dimension-dependent. It could be that (in a given finite dimension) {\it all\/} critical droplets are bounded (we will refer to these simply as ``finite critical droplets''), in which case the distribution of their sizes becomes important in answering fundamental questions of ground state structure and multiplicity (see Sect.~\ref{sec:chaos}). Critical droplets can in principle also be unbounded and contain an infinite subset of spins, but with a boundary that has zero density in the dual lattice to $\mathbb{E}^d$.

There are three models whose critical droplet properties are fully or partially known: the EA model in one dimension, the highly disordered model in all dimensions, and the strongly disordered model in all dimensions. We briefly discuss each below.

\subsection{Critical droplet properties in three models}
\label{subsec:special}

{\it EA model in one dimension.\/} For the EA Ising model on an infinite chain the structure of critical droplets is trivial for the zero-temperature PBC~metastate, which is supported on a single GSP.  For this ground state pair, in any finite window every bond is satisfied, its critical value $J_c=0$, its critical droplet boundary is a single bond (namely itself, or more precisely its dual bond), and its critical droplet is (semi)infinite.  Its flexibility is simply the magnitude of its coupling value, so, e.g., if $\nu$ is Gaussian, then the ground state distribution of flexibilities is a half-Gaussian. 

Although stated for the PBC metastate, these results hold for {\it any\/} metastate generated using coupling-independent boundary conditions (with the proviso that if the boundary conditions are non-spin-flip-symmetric, such as fixed, the resulting metastate is supported on a single ground state). It is instructive to see how the results for the infinite chain follow from a sequence of finite chains with several simple boundary conditions. 

For free BC's, the above results also hold for every {\it finite\/} chain, with $J_c=0$ and the critical droplet extending to a boundary (on either side --- in one dimension only the {\it relative\/} orientation of the spins on either side of the chosen bond matters). As the chain length goes to infinity, the above results are then easily recovered.

The situation is more interesting for periodic, antiperiodic, and fixed BC's. We can divide these into two classes: the first comprises PBC's and fixed BC's with the two boundary spins having the same orientation; the second comprises APBC's and fixed BC's with the two boundary spins having the opposite orientation. For spin chains in the first class, every bond is satisfied if there is an even number of antiferromagnetic couplings; for those in the second class, every bond is satisfied if there is an odd number of antiferromagnetic couplings. The other possibility occurs when chains in the first class have an odd number of antiferromagnetic couplings and chains in the second class have an even number of antiferromagnetic couplings; in that case a single bond must be unsatisfied in the ground state, and this will be the bond having the coupling with smallest magnitude regardless of sign. 

We consider first the case in which there must be one unsatisfied bond in the finite-chain ground state. Denote the bond with smallest coupling magnitude $b_i=\langle x_i,x_{i+1}\rangle$ with coupling value $J_i$. Choose an arbitrary bond $b_0$, different from $b_i$, with coupling value $J_0$ in $J$. Its critical value is then  $J_c={\rm sgn}(J_0)\vert J_i\vert$; when the magnitude of $b_0$'s coupling falls below this value, a droplet consisting of spins lying between an endpoint of $b_0$ and one of $b_i$ will flip. 

The second case is any where {\it all\/} couplings are satisfied in the finite-chain ground state in the initial $J$. Using the same notation as in the first case, as soon as the running coupling value $J(b_0)$ of $b_0$ changes sign, it becomes unsatisfied, but no flip will occur until its magnitude equals $\vert J_i\vert$. (Note that as soon as $J(b_0)$ changes sign, the number of antiferromagnetic couplings changes by one, and now one edge must now be unsatisfied in the ground state.) In this case $J_c=-{\rm sgn}(J_0)\vert J_i\vert$, with all else the same as in the first case.

For both of these cases, as the length of the chain increases, the bond of lowest coupling magnitude moves out to infinity (while randomly switching from the right half to the left half of the chain, but as noted earlier, only the relative orientation of the spins on either side of $b_0$ matters), with its magnitude approaching zero. Therefore as the length of the spin chain increases, the critical value of $b_0$'s coupling decreases to zero, becoming exactly zero in the infinite-chain limit. 

For the EA spin glass on ${\bf Z}^d$, having a finite critical droplet boundary along with an infinite critical droplet is unique to one dimension. The relevant property here is the finiteness of the
critical droplet boundary, rather than the infinite critical droplet ``volume'': as we will see in Sect.~\ref{sec:chaos}, it is the critical droplet boundary, not its volume, that is connected with the presence or absence of ground state multiplicity.  In most respects, though, $d=1$ is a special case and hereafter we focus only on dimensions larger than one.

\medskip 

{\it Highly disordered model in $d$ dimensions.\/}  The highly disordered model~\cite{NS94,NS96a,BCM94} has the EA Hamiltonian~(\ref{eq:EA}) but with a nonphysical, volume-dependent coupling 
distribution (though the couplings remain i.i.d.~for each volume). The coupling distribution is ``stretched'' so that, 
with probability one, in sufficiently large volumes each coupling magnitude occurs on its own scale: it 
is at least twice as large as the next smaller one and no more than half as large as the next
larger one.

A simple way of realizing this is to assign {\it two\/} collections of i.i.d.~random variables $\epsilon_{xy}$ and $K_{xy}$ to the edges, with $\epsilon_{xy}=\pm 1$ each with probability $1/2$, and $K_{xy}$ a continuous random variable
uniformly distributed in the interval~$[0,1]$. We then define the couplings $J_{xy}^{(L)}$ within $\Lambda_L$ as
\begin{equation}
\label{eq:hd}
J_{xy}^{(L)}=c_L\epsilon_{xy}e^{-\lambda^{(L)}K_{xy}}
\end{equation}
where $c_L$ is a scaling factor chosen to ensure a finite energy per unit volume in the thermodynamic limit and $\lambda^{(L)}$ is a scaling parameter that grows quickly
with $L$. It was shown in~\cite{NS96a} that $\lambda^{(L)}\ge L^{2d+1+\delta}$ for any $\delta>0$ is sufficient for the ``stretched'' property described above
to hold in all sufficiently large volumes. Although the couplings themselves depend on $L$, their associated $\epsilon_{xy}$'s and $K_{xy}$'s do not, allowing
ground states and their properties to be well-defined in the infinite-volume limit.

The highly disordered model is one of the few nontrivial spin glass models where the ground state multiplicity is exactly known: it has a single GSP below
six dimensions and uncountably many above~\cite{NS94,NS96a,JR10}. Its critical droplet structure is also known: it was proved (Theorem~2.2 of~\cite{ANS21})
that, assuming there is no percolation at $p_c$ in the corresponding independent bond percolation model, then in all dimensions, the critical droplet 
boundary of any bond in any ground state is finite. Consequently, for any fixed bond in sufficiently large volumes, the critical droplet 
volume is independent of $L$. The {\it distribution\/} of critical droplet sizes, which is relevant for determining ground state multiplicity in 
conventional EA models~(Sect.~\ref{sec:chaos}), is unknown.

{\bf Remark.} As noted above, this result depends on the assumption that there is no percolation at $p_c$ in the
corresponding independent bond percolation model. This has been proved rigorously (see~\cite{FH16} and also Sect.~1.5 of~\cite{Grimmett99}) in all dimensions except $3\le d\le 10$,
but is generally believed to be true in all finite dimensions. 
\medskip

{\it Strongly disordered model in $d$ dimensions.\/}  The strongly disordered model is identical to the highly disordered model except that~(\ref{eq:hd})
is replaced by
\begin{equation}
\label{eq:sd}
J_{xy}=\epsilon_{xy}e^{-\lambda K_{xy}}
\end{equation}
where $\lambda$ (and therefore the coupling distribution) is now independent of $L$. We will be interested in
models where $\lambda\gg 1$.

In the strongly disordered model, the ``stretched'' condition, namely that every coupling value is no more than
half the next larger one and no less than twice the next smaller one, breaks down in sufficiently
large volumes. This can be quantified~\cite{NS96a}: let $g(\lambda)$ be the probability that any two arbitrarily chosen bonds have coupling values that do not
satisfy the highly disordered condition; then $g(\lambda)=2\ln 2/\lambda$. Hence the ``stretched'' condition holds with high likelihood for volumes $\Lambda_L$ in which 
$\vert\Lambda_L\vert^2\ll\lambda$ and likely fails for those with $\vert\Lambda_L\vert^2\gg\lambda$, where $\vert\Lambda_L\vert=L^d$ is the volume (number of spins) of $\Lambda_L$. 

The strongly disordered model is potentially important because its critical droplet
properties are analytically tractable given its similarity to the highly disordered
model. Moreover, since its coupling distribution does not vary with $L$ and is i.i.d.~with mean zero and
finite variance, it satisfies the conditions for $\nu(dJ_{xy})$ given below Eq.~(\ref{eq:EA}). We therefore 
expect global properties such as ground state multiplicity to be the same as in other versions of the EA spin glass with more conventional
coupling distributions.  

Theorem 3.2 of~\cite{ANS21} provides some information on the critical droplet structure of the strongly disordered model:
it shows that, if (as before) there is no percolation at $p_c$ in the corresponding independent bond percolation model, then in the strongly disordered model, the critical droplet boundary of an arbitrary but fixed bond is finite with probability approaching one as $\lambda\to\infty$ (as before, in any ground state in any dimension).

\medskip

We next turn to the connection between critical droplets and disorder chaos, and discuss how critical droplet distributions
determine the size of energy fluctuations (with respect to coupling variations) in ground states.

\section{Incongruent ground states, disorder chaos, and energy fluctuations}
\label{sec:chaos}

Unlike the models discussed in Sect.~\ref{subsec:special}, in the EA model above one dimension we do not yet know the size and/or energy distributions of critical droplets --- which, as we will
see in later sections, would be sufficient to answer the questions posed at the start of the Introduction. In this section we turn to a different question: how are critical droplets related, if at all, to other 
thermodynamic and energetic features of spin glasses?  These questions were addressed in~\cite{ANS19}, and this section provides a brief review of results from that paper relevant to the current discussion.  The main results of interest are contained in theorems which concern the relation between critical droplets and disorder chaos, and which show how critical droplet properties determine energy fluctuations (with respect to coupling variations) in the ground state. Knowledge of the latter in a given dimension could enable a determination of ground state multiplicity in that dimension, as we will see.

We begin by discussing disorder chaos, which was introduced as a feature of the droplet-scaling picture~\cite{BM87,FH88b}, but has since been proved to occur~\cite{Chatterjee09} also in the infinite-range SK~model~\cite{SK75}, and so may be a feature of {\it all\/} pictures of the spin glass phase in short-range models. In the context of the EA Ising model it has not been proved to occur (or not occur) in any dimension greater than one; the only rigorous result in this context we are aware of is a weak bound proved by Chatterjee~\cite{Chatterjee09} on the {\it amount\/}  of disorder chaos that can occur in the EA model. 

Disorder chaos was originally defined as follows (here we follow the discussion of~\cite{KB05}): consider the EA~Hamiltonian~(\ref{eq:EA}) with Gaussian $\nu(dJ_{xy})$ (with mean zero and variance one, as always). One wants to perturb {\it every\/} coupling by a small random amount such that the modified couplings still obey the law of the original~$\nu(dJ_{xy})$. One way of doing this (another will be discussed below) is to introduce a small~$\Delta J>0$ and then modify the couplings by
\begin{equation}
\label{eq:mod1}
J_{xy}\, \rightarrow\, J_{xy}'=\frac{J_{xy}+\eta_{xy}\Delta J}{\sqrt{1+(\Delta J)^2}}\, ,
\end{equation}
where $\eta_{xy}$ is an i.i.d.~random variable also taken from a Gaussian distribution with mean zero and variance one.  In a fixed volume with specified BC's, one then compares, using various overlap functions, the ground state for the original $J$ with that for~$J'$.  According to the droplet-scaling picture, there is a critical length $\ell_c$, depending on $\Delta J$, such that on lengthscales small compared to~$\ell_c$ the spin overlap of the primed and unprimed ground states is close to one, while for lengthscales large compared to $\ell_c$ the spin overlap falls off rapidly as lengthscale increases. In the droplet-scaling picture the dependence of $\ell_c$ on~$\Delta J$ is $\ell_c=(\Delta J)^{-1/\xi}$ with $\xi$ an exponent that can be related to others that arise naturally in droplet-scaling. 

The effects of the coupling transformation~(\ref{eq:mod1}) on the EA model are simple to work out in one dimension for a given realization of couplings and $\eta_{xy}$'s.  Select any arbitrary site as the origin. A coupling will become unsatisfied, thereby flipping a critical droplet, when $\Delta J$ is sufficiently large so that its $J'_{xy}$ has opposite sign to its $J_{xy}$ (of course this can only happen if its $J_{xy}$ and its $\eta_{xy}$ have opposite sign). Therefore, as $\Delta J$ decreases to 0 the coupling nearest to the origin which first changes sign moves out to infinity (whether such a coupling is to the right or left of the origin changes randomly as $\Delta J$ decreases). As a result the ground state pair remains unchanged on an increasing lengthscale $\ell_c\to\infty$ as $\Delta J\to 0$.

Before turning to the results for disorder chaos in~\cite{ANS19}, a brief digression is helpful. In~\cite{ANS19} (as well as~\cite{Chatterjee09}) the focus is on edge, rather than spin, overlaps (though it should be noted that the function~$C(r,L)$ defined in Eq.~(3) of~\cite{KB05} reduces to an edge overlap for $r=1$).
This is because a central question in~\cite{ANS19}, as in this paper, is whether {\it incongruent ground states\/} exist in the EA model. The term ``incongruent'' as applied to spin glasses first appeared in~\cite{HF87,FH87}, and (in the context of ground states) refers to two infinite-volume ground states whose edge overlap is strictly smaller than one; that is, there is a positive fraction of edges that are satisfied in one ground state and unsatisfied in the other, and vice-versa. More precisely, let
\begin{equation}
\label{eq:edge}
Q_L(\sigma,\sigma')=\frac{1}{\vert E_L\vert}\sum_{\langle x,y\rangle\in E_L}\sigma_x\sigma_y\sigma'_x\sigma'_y
\end{equation}
denote the edge overlap between (any) two spin configurations $\sigma$ and $\sigma'$ both in $\Lambda_L$, with $\vert E_L\vert$ denoting the total number of edges (including those touching the boundary) contained within~$\Lambda_L$. Then two infinite-volume GSP's $\eta$ and $\zeta$ are incongruent if $\lim_{L\to\infty} Q_L(\eta_L,\zeta_L) = Q(\eta,\zeta)<1$, where $\eta_L$ is the restriction of $\eta$ to $\Lambda_L$ and similarly for $\zeta_L$. (More precisely this limit is a lim~sup and a coupling-dependent subsequence of volumes may need to be taken, but that and other technical issues are discussed in~\cite{ANS19} and need not concern us further here.)

The reason for this digression is that all mentions of multiple ground states throughout this paper refer strictly to incongruent states, which we have argued elsewhere~\cite{NS01c,NS02,NS03b,NS06a} are the ``physical'' states; multiple spin glass states that are not incongruent are of mathematical interest but unlikely to appear in any laboratory or other physical setting. In particular, both many-state pictures discussed in Sect.~\ref{sec:interfaces} below are limited to incongruent states. In fact, it was proved in complete generality~\cite{NS01c} (i.e., independent of any particular picture) that {\it any\/} zero-temperature metastate generated by coupling-independent BC's is supported on either a single ground state (or GSP) or else on multiple {\it incongruent\/} ground states (or GSP's). We will discuss this further in Sect.~\ref{sec:interfaces}, but now return to disorder chaos as defined in~\cite{ANS19}.

In~\cite{ANS19} the notion of disorder chaos is extended to specify not only {\it if\/} disorder chaos occurs, but also the {\it scale\/} at which it occurs. To do this the coupling perturbation is redefined to a form more convenient for our purposes: let $J$ and $J'$ denote two independent realizations of the couplings, both drawn from a Gaussian~$\nu$ with mean zero and variance one. Then for a particular edge~$\langle x,y\rangle$, if $J_{xy}$ is drawn from $J$ and $J_{xy}'$ from $J'$, then we consider an interpolation $J_{xy}(t)$ between $J_{xy}$ and $J'_{xy}$ parametrized by $t\ge 0$:
\begin{equation}
\label{eq:mod2}
J_{xy}(t)=e^{-t}J_{xy}+\sqrt{1-e^{-2t}}J'_{xy}\, .
\end{equation} 
Unlike most other treatments, the setting in~\cite{ANS19} was confined to infinite-volume ground states, so the perturbation~(\ref{eq:mod2}) is applied only within a fixed volume~$\Lambda_L\subset\mathbb{Z}^d$, with $J_{xy}(t)=J_{xy}$ outside~$\Lambda_L$. Note that if we identify $t$ with $\frac{1}{2}(\Delta J)^2$ and take the limit $t\to 0$, the perturbations~(\ref{eq:mod1}) and~(\ref{eq:mod2}) converge for sufficiently small $t$. 

Using the interpolation formula~(\ref{eq:mod2}) allows a straightforward extension of the idea of disorder chaos. Let $\sigma_L(0)$ denote the periodic boundary condition ground state pair in~$\Lambda_L$ with realization~$J$ and $\sigma_L(t)$ denote the same with realization~$J(t)$. We can now state the following (informal) definition~(for the full definition, see Def.~1.3 of~\cite{ANS19}): we will say there is {\it absence\/} of disorder chaos at scale~$\alpha$, with $0\le\alpha\le 1$, if with probability close to one the edge overlap $Q_L\bigl(\sigma_L(0),\sigma_L(t)\bigr)$ remains close to one for~$t\le C\vert\Lambda_L\vert^{-\alpha}$, with $C$ a constant, for all sufficiently large~$L$.  Qualitatively speaking, disorder chaos is completely absent when $\alpha=0$, while $\alpha=1$ corresponds to what could reasonably be called ``strong'' disorder chaos. Between these limits, the amount of disorder chaos can effectively be ``tuned'' be varying alpha from 0 to 1 (in a purely theoretical sense; the quantity $\alpha$ is fixed within a given system or model and plays an important role in its thermodynamics, as we will discuss below).

One of the main results of~\cite{ANS19} was to find a relation between the amount of disorder chaos in a system and the size of its critical droplets; we emphasize again that by ``size'' of a critical droplet we refer to the number of edges in its boundary $\vert\partial D(b_{xy},\sigma_L)\vert$ and not to the number of spins contained within the droplet itself. As before fix a volume $\Lambda_L$ with GSP $\sigma_L$ and choose an arbitrary bond $b=\langle x,y\rangle$ inside~$\Lambda_L$. Let $0<C<\infty$ be a constant (i.e., independent of $L$). Then Theorem~1.5 of~\cite{ANS19} states that

\medskip

{\bf Theorem~4.1~(Arguin-Newman-Stein~\cite{ANS19}).} Suppose there exists $\gamma$ with $0\le\gamma\le 1$ such that with probability one, for all large $\Lambda_L$,
\begin{equation}
\label{eq:cddc}
\vert\partial D(b,\sigma_L)\vert\le C\vert\Lambda_L\vert^\gamma\,\,\,\,\, {\rm for}\,\,\, {\rm all}\,\,\,\,\, b\in\mathbb{E}_L\, .
\end{equation}
Then there is absence of disorder chaos on all scales $\alpha>2\gamma$.

\medskip

{\bf Remark.} While the proof of Theorem~1.5 of~\cite{ANS19} is technical, the underlying idea is to show that if all critical droplets are limited in size by a specified amount, and the couplings are globally perturbed by a sufficiently small amount, many of the edge flexibilities will not decrease to zero, and thus the resulting ground state will agree with the original on most edges. The result can be controlled precisely enough so that the ``tuning'' of $\gamma$ determines the $\alpha$ above which disorder chaos does not occur.

\medskip

So disorder chaos and critical droplets are different but related facets of ground state stability in spin glasses, and therefore of spin glass thermodynamics. While disorder chaos by itself is an interesting phenomenon, especially given its conjectured relation with temperature chaos~\cite{BM87,FH88b}, it also plays a primary role in determining ground state multiplicity, as we will now see. 

The connection between disorder chaos and ground state multiplicity is that the amount of disorder chaos in a spin glass controls the {\it energy fluctuations\/} in the ground state, as determined in Corollary~1.4 of~\cite{ANS19}. Here one assumes that two or more incongruent ground states (or GSP's) exist, and chooses an arbitrary incongruent pair~$\sigma^1$ and $\sigma^2$. Corollary~1.4 (here stated as a theorem) then says the following:

\medskip

{\bf Theorem~4.2~(Arguin-Newman-Stein~\cite{ANS19}.)} If there is absence of disorder chaos at scale~$\alpha$~($0\le\alpha\le 1$), then given two (infinite-volume) incongruent ground states (or GSP's) $\sigma^1$ and $\sigma^2$, there exists $C>0$ independent of~$L$ such that
\begin{equation}
\label{eq:flucs1}
{\rm Var}\bigl({\cal H}_{L,J}(\sigma^1)-{\cal H}_{L,J}(\sigma^2)\bigr)\ge C\vert\Lambda_L\vert^{1-\alpha}
\end{equation}
where ${\cal H}_{L,J}$ denotes the EA Hamiltonian~(\ref{eq:EA}) restricted to the volume $\Lambda_L$ and the variance is taken with
respect to all the couplings inside~$\Lambda_L$ (i.e., in $\mathbb{E}_L$).

\medskip

The connection to ground state multiplicity arises because of a (rigorous) finite-volume bound on free energy fluctuations found independently in~\cite{AFunpub} and~\cite{NSunpub} (the argument can be found in~\cite{Stein16}). Fix~$J$ and consider two {\it finite-volume\/} ground states $\sigma_L^P$ and $\sigma_L^{AP}$ generated in a volume $\Lambda_L$ at fixed $T\ge 0$ using periodic boundary conditions ($\sigma_L^P$ ) and antiperiodic boundary conditions ($\sigma_L^{AP}$). Their free energy difference is $\Delta F_L= F(\sigma_L^P)-F(\sigma_L^{AP})$. Then 
\begin{equation}
\label{eq:oldflucs}
{\rm Var}(\Delta F_L)\le AL^{d-1}
\end{equation}
with $A$ a constant and the variance as before is over all couplings in~$\mathbb{E}_L$. At zero temperature the lefthand side of~(\ref{eq:oldflucs}) reduces to the lefthand side of~(\ref{eq:flucs1}), i.e., it represents energy fluctuations between ground states within a fixed volume.

The result~(\ref{eq:oldflucs}) is equivalent to stating that the energy difference between incongruent ground states in volumes~$\Lambda_L$ cannot increase faster than $O(L^{(d-1)/2})$.
This would violate the lower bound~(\ref{eq:flucs1}) if $\alpha<1/d$ in dimension~$d$; if this were to happen, incongruent ground states would be ruled out in that dimension. (This bound on $\alpha$ appeared in Claim~5.4 in~\cite{ANS19}.) 

It should be noted that such a result would not yet be rigorous because the lower bound~(\ref{eq:flucs1}) was derived for infinite-volume ground states restricted to a volume $\Lambda_L$ while the upper bound~(\ref{eq:oldflucs}) was derived for finite-volume ground states in $\Lambda_L$. Nonetheless, if the lower bound~(\ref{eq:flucs1}) were to hold for the infinite-volume case restricted to finite volume, it seems likely to hold in the pure finite-volume case as well.

Because of the relation~(\ref{eq:cddc}), Theorem~4.2 can be recast as a connection between critical droplets and energy fluctuations:

\medskip

{\bf Theorem~4.3.} Let $\alpha$ denote the minimal scale on which disorder chaos is absent and $\gamma$ be defined as in~(\ref{eq:cddc}). Then Theorems~4.1 and~4.2 together imply
\begin{equation}
\label{eq:flucs2}
{\rm Var}\bigl({\cal H}_{L,J}(\sigma^1)-{\cal H}_{L,J}(\sigma^2)\bigr)\ge C\vert\Lambda_L\vert^{1-2\gamma}\, .
\end{equation}

\medskip

{\bf Remark.} If the finite-volume upper bound~(\ref{eq:oldflucs}) were to hold for the infinite-volume case restricted to finite volume, then a sufficient condition for absence of incongruence is $\gamma<(2d)^{-1}$.

We can test this result for the EA model in one dimension. As noted earlier, the critical droplet boundary in that case consists of a single bond, so $\gamma=0$, satisfying the condition for absence of incongruence (as well as the absence of disorder chaos on any scale). As is well known, there is only a single GSP for the EA model in one dimension, consistent with the result obtained above.

In higher dimensions the behavior of critical droplets sets bounds both on the scale of disorder chaos and on the magnitude of energy difference fluctuations between (potential) incongruent states. Perhaps the simplest possibility is that all critical droplets are finite with an exponential size distribution. If that were to occur in some dimension, $\gamma=0$ and it becomes very likely that the PBC metastate is supported on only a single GSP in that dimension.  Other (reasonable) possibilities are that all critical droplets are finite but with a power-law falloff (with the power possibly dimension-dependent); or it could be that in a given dimension some positive density of edges have infinite critical droplets, which is a particularly interesting case. We will return to these issues in Sect.~\ref{sec:rsb}.

\section{Interfaces and low-temperature spin glass scenarios}
\label{sec:interfaces}

Many pictures have been proposed for the thermodynamics of the low-temperature phase of the EA model. We focus here on four of these, of which two 
(replica symmetry breaking, or RSB~\cite{Read14,Parisi79,Parisi83,MPSTV84a,MPSTV84b,MPV87,MPRRZ00}, and scaling-droplet~\cite{Mac84,BM85,FH88b,FH86}) have already been referred to in previous sections, given that they have  received a great amount of attention in the literature.  The other two are the so-called trivial-nontrivial (TNT) picture~\cite{PY00,KM00} and the chaotic pairs (CP) picture~\cite{NS96b,NS97,NS98}. The reason for considering these four pictures together is encapsulated in Table~1, as we now discuss.

Of the four, two (RSB and chaotic pairs) predict the existence of many ground states, and the other two (scaling-droplet and TNT)
imply the existence of only a single pair of spin-reversed ground states~\cite{FH88b,NS01c,ANS19}.  These four pictures are characterized not only by the multiplicity of ground states they predict (which doesn't distinguish between the two columns in Table~1, i.e., the ground state multiplicity is the same across either row), but more fundamentally by the nature of the {\it interfaces\/} that separate their ground states from their lowest-lying excitations. In fact, the point of view of this paper is that the {\it more fundamental property of the differing spin glass scenarios is the geometric and energetic nature of the interfaces separating ground states from their lowest-energy large-lengthscale excitations\/}. The presence or absence of multiplicity of ground states {\it follows\/} as a consequence of the nature of these excitations, and so in this sense is less fundamental.

An interface between any two infinite spin configurations (not necessarily ground states) $\eta$ and $\zeta$ is defined to be the set of bonds whose associated couplings are satisfied in $\eta$ and unsatisfied in $\zeta$, or vice-versa; they separate regions in which the spins in $\eta$ agree with those in $\zeta$ from regions in which their spins disagree. An interface may consist of a single connected component or multiple disjoint ones (often referred to as {\it domain walls\/}). If $\eta$ and $\zeta$ {\it are\/} ground states, then given~(\ref{eq:gs}), any component of their mutual interface must be infinite in extent.  Note that a GSP $(\eta,\overline{\eta})$, in which $\eta$ and $\overline{\eta}$ are simply global flips of each other, have no interface between them, and as noted earlier we will affix a single label~($\eta$ in this case) to the pair. 

Interfaces can be characterized by their geometry and energy. With respect to the former, they can be either ``space-filling'', meaning they comprise a positive density of all bonds in~$\mathbb{E}^d$, or zero-density --- i.e., the dimensionality of the interface in a $d$-dimensional system is strictly less than $d$.  (Recall that two incongruent ground states, by definition, differ by a space-filling interface.) Moreover, their energies can scale with distance along an interface: as one travels a distance $\ell$ along an interface between $\eta$ and $\zeta$, the magnitude of its energy change (which is equal in $\eta$ and $\zeta$; only the sign differs) might scale as $\ell^y$, $y>0$, or else might remain $O(1)$, independent of length traveled along the interface ($y=0$). For our purposes it is sufficient to distinguish only between an interface whose energy increases with $\ell$, which we'll denote a ``high-energy interface'', and one whose energy remains $O(1)$ no matter how far one travels along the interface (a ``low-energy interface''). For a high-energy interface, it is not required that the interface energy depends on $\ell$ as a power law; all that matters is that the energy increases (not necessarily monotonically) without bound as $\ell\to\infty$.

With this in mind we present the four low-temperature spin glass scenarios in the following table, which illustrates their various relationships (and clarifies why we consider these four pictures together):


\bigskip

\begin{center}
\begin{tabular}{ r|c|c| }
\multicolumn{1}{r}{}
 &  \multicolumn{1}{c}{Low-energy}
 & \multicolumn{1}{c}{High-energy} \\
\cline{2-3}
Space-filling&RSB&Chaotic pairs \\
\cline{2-3}
 Zero-density&TNT&Scaling-droplet\\
\cline{2-3}
\end{tabular}

\medskip

\begin{singlespace}
\small{{\bf Table~1.} The four scenarios described in the text for the low-temperature phase of the EA model, categorized in terms of interface geometry (rows) and energetics (columns). At zero temperature the vertical column headings describe the energy of the minimal large lengthscale excitations above the ground state predicted by each; at low temperature they describe the free energy of these excitations. Adapted from Fig.~1 of~\cite{NS03a}.}
\end{singlespace}

\bigskip

\end{center}

{\bf Remark on Table~1.} The scaling-droplet picture predicts a broad distribution of (free) energies for a minimal energy compact droplet of diameter~$O(L)$, with a characteristic energy growing as $L^\theta$ with $\theta>0$ in dimensions where a low-temperature spin glass phase is present. The distribution is sufficiently broad that there exist droplets of~$O(1)$~energy on large lengthscales, but these appear with a probability varying as $L^{-\theta}$ as $L\to\infty$. In contrast, both the RSB and TNT pictures require droplets with $O(1)$ energy to appear with positive probability bounded away from zero on all lengthscales. For this reason the scaling-droplet scenario belongs in the second column of Table~1.

\medskip

As proved elsewhere~\cite{NS02}, the presence of space-filling excitations, regardless of energetics (as long as the energy scales no faster than $O(L^{(d-1)/2})$, as discussed below~(\ref{eq:oldflucs})), is a sufficient condition for ground state multiplicity, so the scenarios in the first row (RSB and chaotic pairs) both imply the presence of multiple GSP's.  Their difference in interface energetics, however, already shows that the two pictures are {\it not\/} equivalent at zero temperature, as sometimes claimed. In fact, the interface energetics leads to the most significant difference between the two pictures at low but nonzero temperature, in which a single thermodynamic state in the RSB picture comprises a nontrivial mixture of infinitely many incongruent pure state pairs with free energies differing by $O(1)$~\cite{MPV85,DT85}, while in chaotic pairs a single thermodynamic state consists of a single {\it pure state\/} pair. Therefore, in finite volumes the two pictures appear very different, given that the properties of infinite-volume thermodynamic states are reflected in those of the finite-volume Gibbs states observed in fixed windows as described in Sect.~\ref{subsec:pbc}. The (low temperature) PBC metastate in both pictures is supported on an infinite (uncountable in the case of RSB~\cite{NS06b}) set of distinct thermodynamic states, but as noted, these are each mixtures of incongruent pure states in RSB but not in chaotic pairs.

As an interesting side note, a rigorous analysis~\cite{ANS15} of pure state weight distributions in potential mixed-state pictures strongly suggests that RSB is the {\it only\/} viable mixed-state picture that can be supported in the EA model.

So the two pictures in the top row of Table~1 can be shown to imply the presence of multiple ground state pairs in the support of $\kappa_J$. What about the two in the bottom row? A proof that the TNT picture, as put forward in~\cite{PY00,KM00}, implies a single ground state pair appears in~\cite{NS01c}.  The caveat is important, however. The procedures used in~\cite{PY00,KM00} will be discussed in more detail in~Sect.~\ref{sec:rsb}; here we need note only that both procedures, when adapted to the metastate approach, use in the infinite-volume limit a translation-invariant method to generate zero-density, $O(1)$-energy large-lengthscale excitations above the ground state in large volumes. The translation-invariance implies that the excitations deflect to infinity as the volume size increases (see~\cite{NS01c} for details) resulting in a two-state picture. This leaves open the possibility of generating infinite, zero-density, $O(1)$-energy excitations in a {\it non\/}-translation-invariant way, which might or might not lead to a many-state picture. 

Turning to droplet-scaling, an argument that this picture also implies a single ground state pair appears in~\cite{ANS19}; however, it relied on the conjecture that the upper bound~(\ref{eq:oldflucs}) holds for the infinite-volume case restricted to finite volume, as discussed above Theorem~4.3. If that conjecture is proved correct then the argument becomes fully rigorous. 

It is then reasonable to conclude that while the pictures in the first row predict multiple ground states in the support of $\kappa_J$, those in the second predict a single spin-reversed pair. This is the primary reason behind our assertion above that interface properties are more fundamental than ground state multiplicity.

\section{Space-filling critical droplets and $\sigma$-criticality}
\label{sec:sfcd}

Up until now we have collected a diverse series of previously published results that together argue for the view that critical droplets play a fundamental role in
determining ground state properties in spin glasses. In particular we have seen that they are intimately connected to and determine ground state properties such as disorder chaos and
ground state multiplicity. Moreover, the material already discussed provides the background and terminology required to present new results that further this argument, which will include establishing the connection between critical droplet distributions and the low-temperature scenarios listed in Table~1. 

As argued in previous sections, a ground state can be characterized by its distribution of critical droplet geometries and energies (or equivalently, edge flexibilities). As discussed in previous sections, critical droplets can be finite or infinite, and if the latter, can be either space-filling or zero-density. It is important to note that, unlike interfaces, critical droplets can {\it only\/} have energies of~$O(1)$ no matter their size, given~(\ref{eq:cd}) and the inequality~(\ref{eq:flexbound}). 

It is not hard to show that a positive fraction of bonds must have finite critical droplets in any ground state in any dimension, but we omit the proof here.  A natural division is then  between a ground state having all its critical droplets finite vs.~having a positive fraction of its critical droplets infinite with the remainder finite. Infinite critical droplets can be further subdivided into those which are space-filling and those which are zero-density; recall from previous sections that when we refer to the ``size'' of a critical droplet, we are referring not to the number of spins it contains but rather the number of edges whose duals lie in its boundary.

This categorization, while useful, isn't comprehensive, because it considers only the critical droplets assigned to each edge. We will see below that it is equally important to consider for each edge how many distinct critical droplets (i.e., those assigned to {\it other\/} edges) to which it belongs. (When we say an edge ``belongs'' to a critical droplet, we mean that its dual lies on the boundary of another edge's critical droplet.)  We therefore introduce a new definition that extends the concept of a space-filling critical droplet and which will be useful in what follows.

\medskip

{\bf Definition 6.1.}  Choose a coupling realization $J$ and an arbitrary bond $b_{xy}=\langle x,y\rangle$, and consider a ground state $\sigma$ consistent with $J$. Then $b_{xy}$ will be called $\sigma$-{\it critical\/} in ground state $\sigma$ if there is an open interval of coupling values $J(b_{xy})$ throughout which $\sigma$ remains a ground state and $b_{xy}$ belongs to the critical droplet boundary $\partial D_{b_{x'y'},\sigma}$ of a positive density of bonds $b_{x'y'}\in\mathbb{E}^d$.

\medskip

Before proceeding we will need the following lemma:

\medskip

{\bf Lemma 6.2.} Suppose a bond $b_1$ with coupling value $J_1$ in $J$ and critical value $J_c$ in $\sigma$ belongs to the critical droplet boundary $\partial D(b_2,\sigma)$ of a different bond $b_2$. Then $b_1$ will remain in $\partial D(b_2,\sigma)$ for the entire range of coupling values between $J_1$ and $J_c$.

\medskip

{\bf Proof.}  If $b_1\in\partial D(b_2,\sigma)$ when $J(b_1)=J_1$ then for this coupling value $D(b_2,\sigma)$ is the minimum energy droplet to which $b_2$ belongs. Without loss of generality assume $J_1>J_c$. Then when $J(b_1)$ is lowered to a value $J^*\in(J_c,J_1]$, the energy of $D(b_2,\sigma$) is lowered by an amount $J_1-J^*$. Then by Lemmas~2.6 and~3.1 the energy of every other droplet (critical or otherwise) in $\sigma$ is either lowered by the same amount (if its boundary includes the dual of $b_1$) or else is unaffected. Consequently $D(b_2,\sigma$) remains the lowest-energy droplet in $\sigma$ which passes through $b_2$, and the result follows.

\bigskip

In what follows it will be important to distinguish between two types of $\sigma$-criticality. We will say a bond exhibits $\sigma${\it -criticality of the first kind\/} if its critical droplet is space-filling. A bond exhibits $\sigma${\it -criticality of the second kind\/} if its critical droplet boundary is not space-filling but if, for some open interval of coupling values (extending down to its critical value), it nevertheless belongs to the critical droplet boundary of a positive density of edges in ${\mathbb E}^d$. 

We will see in Theorem~6.3 below that if a bond's critical droplet is space-filling, then it has a nonzero range of coupling values for which a positive density of edges shares the {\it same\/} critical droplet --- namely that of the original bond. In contrast, the second kind of $\sigma$-criticality occurs when a coupling belongs to an infinite, positive-density set of {\it different\/} critical droplets. This could in principle occur in situations wherein a positive density of bonds has infinite but zero-density critical droplets, or even a situation wherein all critical droplets are finite but with a sufficiently slow falloff of critical droplet sizes (this will be explored further in Sect.~\ref{sec:multi}).  

What is shared in both kinds of $\sigma$-criticality is that, within a certain range of coupling values, altering the coupling value of a $\sigma$-critical edge by a small amount (i.e., without causing a droplet flip) changes the flexibilities of a positive density of bonds in $\sigma$; when this occurs we will say that such a bond {\it controls\/} the flexibilities of the affected bonds. For the second kind of $\sigma$-criticality this is true by definition.  As for a bond which exhibits $\sigma$-criticality of the first kind, the next theorem shows that it has a nonzero range of coupling values in which it too controls the flexibilities of a positive density of bonds in $\sigma$. 

\medskip

{\bf Theorem 6.3.}  There is an open interval of coupling values, with the critical value at one end of the interval, for which a bond which is $\sigma$-critical of the first kind controls the flexibilities of a positive density of bonds without causing a droplet flip in~$\sigma$.

\medskip

{\bf Proof.}  Given a coupling realization~$J$ and a corresponding ground state~$\sigma$, let $b_0$ be a $\sigma$-critical bond of the first kind, with coupling value $J_0$ in $J$. For ease of discussion, and without loss of generality, let $J_0$ be positive and satisfied in~$\sigma$, and denote its critical value in $\sigma$ by $J_c$; then $J_0>J_c$ and the flexibility of $b_0$ in $\sigma$ is $f(b_0,\sigma)=J_0-J_c$.  

Recall from Lemmas~2.6 and~3.1 that lowering the coupling value of $b_0$ to any value in the interval $(J_c,J_0)$ cannot increase the flexibility of any other coupling in $\sigma$. With this in mind, consider a different bond $\tilde b_j\in\partial D(b_0,\sigma)$. For any such bond, its critical droplet will also be $D(b_0,\sigma)$ unless it belongs to a closed surface~$S_j$ (as always, in the dual lattice), different from $\partial D(b_0,\sigma)$ (though there may be bonds common to both), with $E(S_j)<E\bigl(D(b_0,\sigma)\bigr)$ (if there exists more than one such surface, then the $S_j$ with the smallest $E(S_j)$ is $\partial D({\tilde b}_j,\sigma)$). 
As $J(b_0)$ is lowered from $J_0$ toward $J_c$, $E\bigl(D(b_0,\sigma)\bigr)$ is simultaneously lowered while $\sigma$ remains unchanged. During this process, as long as the inequality $E\bigl(D(b_0,\sigma)\bigr)>E(S_j)$ remains satisfied, $E(S_j)$ remains unchanged by Lemma~2.5(a). But because $E(S_j)>0$ and $E\bigl(D(b_0,\sigma)\bigr)=0$ exactly at $J(b_0)=J_c$, eventually $E(S_j)>E\bigl(D(b_0,\sigma)\bigr)$ before $J_c$ is reached.

We now make the following claim: as $J(b_0)$ is lowered toward $J_c$, it will eventually arrive at a value $J^*\in(J_c,J_0]$ such that a positive density of couplings in $\partial D(b_0,\sigma)$ (and therefore a positive density of couplings in $\mathbb{E}^d$)  will have the same critical droplet as $b_0$.  To see this, set $J(b_0)=J_c$ (note that the probability of this occurring with $J(b_0)$ chosen from a continuous distribution is zero). In that case $E\bigl(D(b_0,\sigma)\bigr)=0$, and because in the original $J$ the flexibility of all bonds was strictly positive, every bond $b_j\in\partial D(b_0,\sigma)$ now has flexibility zero; by Lemma~2.5(b) all now share the critical droplet $D(b_0,\sigma)$. Now take $J(b_0)=J_c+\epsilon$ for some $\epsilon>0$. If the above claim is false, then for {\it any\/} $\epsilon>0$ a fraction one of all bonds $b_j\in\partial D(b_0,\sigma)$ must have a critical droplet $D(b_j,\sigma)$, different from $D(b_0,\sigma)$, with  $E\bigl(D(b_j,\sigma)\bigr)<\epsilon$. But this can only be true if $E\bigl(D(b_j,\sigma)\bigr)=0$ for all but a zero-density set of $b_j$, which cannot happen given that couplings are chosen from a continuous distribution, as noted above. This proves the claim.

\bigskip

It is important to recognize that {\it no\/} bond can ever control the flexibilities of other bonds over its entire range of coupling values. In particular a bond cannot belong to the critical droplet boundary of another bond if its coupling value is such that is ``supersatisfied''~\cite{NS2D00}, i.e., necessarily satisfied in every GSP.   Upper and lower bounds for this range are easily derived, and are as in~(\ref{eq:flexbound}): a sufficient condition for a bond $b_{xy}$ to be supersatisfied is if its coupling value $J_{xy}$ satisfies
\begin{equation}
\label{eq:ss}
\vert J_{xy}\vert\ge J_{\rm max}={\rm min}(\sum_{z\ne y\atop |z-x|=1}\vert J_{xz}\vert,\sum_{u\ne x\atop |y-u|=1}\vert J_{uy}\vert)\, .
\end{equation}

The following lemma will be useful later:

\medskip

{\bf Lemma 6.4.} There is a nonzero gap between the bounds in~(\ref{eq:ss}) and the critical value $J_c$ in any ground state of any edge that is $\sigma$-critical in that ground state; i.e., $-J_{\rm max}<J_c<J_{\rm max}$.

\medskip

{\bf Proof.} For a given bond $b_{xy}$, its critical value must lie in the interval~$[-J_{\rm max},J_{\rm max}]$ given by~(\ref{eq:ss}); moreover, its critical value can only attain the values at the endpoints of this interval if its critical droplet ---  call it $D_{\rm max}$ --- is the same as that used to derive the bounds. These bounds arise from the droplet flip of a single spin, which is at an endpoint of the edge~$\langle x,y\rangle$ in question; the droplet boundary is also local, comprising the duals of $\langle x,y\rangle$ and the edges in the smaller sum on the right-hand side. If $\langle x,y\rangle$ is a $\sigma$-critical edge of the first kind in~$\sigma$, its critical droplet is space-filling, and therefore different from $D_{\rm max}$.  This proves the result for a critical droplet of the first kind.

If~$\langle x,y\rangle$ is a $\sigma$-critical edge of the second kind in~$\sigma$, it belongs to the critical droplet boundary of an infinite number of other edges. By Lemma~2.5(a), its flexibility --- and hence the energy of its critical droplet --- can be no greater than that of any of the critical droplets to whose boundaries it belongs. Therefore its flexibility must be less than or equal to the minimum (or infimum) of the energy of this infinite collection.  All of these energies (but one, only if $D_{\rm max}$ is the critical droplet of one of the neighboring bonds of $b_{xy}$) must be strictly less than $D_{\rm max}$. Therefore the energy of the critical droplet of $b_{xy}$ is strictly smaller than $E(D_{\rm max})$, from which the statement of the theorem follows.

\bigskip

We will see in Sect.~\ref{sec:multi} that if $\kappa_J$ is supported on a single GSP, then there is zero density of bonds being $\sigma$-critical of either kind. An extreme case is one where all critical droplets in a given dimension are finite with an exponential distribution of sizes. This is consistent with the argument in~Sect.~\ref{sec:chaos} that if a spin glass ground state has an exponential distribution of critical droplet sizes, then energy fluctuations will be sufficiently large so that ground state multiplicity should be absent. 

In what follows we shall refer to a bond being $\sigma$-critical without further specification if it makes no difference whether the $\sigma$-criticality is of the first or second kind.

\section{Some consequences of $\sigma$-criticality}
\label{sec:consequences}

In the EA model at positive temperature, if thermodynamic states consist of a nontrivial mixture of incongruent pure state pairs (as in the RSB picture), then two things must be true: the number of pure state pairs in any single thermodynamic state is (countably or uncountably) infinite~\cite{NS10}, and the PBC metastate at that temperature is supported on an uncountable infinity of pure state pairs~\cite{NS06b}. Until now there have been no equivalent rigorous statements for zero temperature. With the machinery developed in previous sections, we are now able to make some progress on these questions; in particular, we will show in this and succeeding sections that the zero-temperature PBC metastate $\kappa_J$ in the RSB picture is supported on at least a countable infinity of pure states.

One obstacle in extending positive-temperature results to zero temperature has been the lack of a translation-invariant measure on ground state pairs that applies in all possible scenarios.
That can now be remedied with the following definition:

\medskip

{\bf Definition 7.1.}  For fixed $J$, let $P_J(f,\sigma)$ denote the (empirical) probability distribution over all edges of the flexibilities $f$ in the ground state $\sigma$, and $P_J(f)=\langle P_J(f,\sigma)\rangle_{\kappa_J}$ be the metastate average of $P_J(f,\sigma)$ over the ground states~$\sigma$ in the support of $\kappa_J$.

\medskip

It is easy to see that $P_J(f)$ is both measurable and translation-invariant. This leads to the following lemma, which will be useful in what follows:

\medskip

{\bf Lemma 7.2.} $P_J(f)$ is almost surely~constant (i.e., constant except for a set of measure zero) with respect to $J$.

\medskip

{\bf Proof.\/} The i.i.d.~coupling distribution~$\nu(J)$ is translation-ergodic and $P_J(f)$ is a measurable, translation-invariant function on the $J$'s. The result immediately follows.

\bigskip

As a consequence of Lemma~7.2, we can drop the subscript and simply write $P(f)$ for the metastate-averaged flexibility distribution. 

\subsection{Dependence of $P_J(f,\sigma)$ on~$\sigma$}
\label{subsec:dependence}

One possible use of $P_J(f,\sigma)$ is to provide a (translation-invariant) way of distinguishing among ground states. The following theorem shows that it does,
at least in cases where ground states have a positive fraction of $\sigma$-critical edges.  
 
\medskip

{\bf Theorem~7.3.}  Suppose that $\kappa_J$ is supported on multiple incongruent GSP's $\sigma$, and suppose that a subset of $\sigma$'s having overall positive weight in $\kappa_J$ each have a positive fraction of $\sigma$-critical edges. Then $P_J(f,\sigma)$ must have some dependence on $\sigma$.

\medskip

{\bf Proof.}  From the assumption of the theorem, there is a set of GSP's $\sigma$ with positive weight in $\kappa_J$ with the property that $\sigma$ has a positive density of $\sigma$-critical edges. Let $b_0$ be an edge that is $\sigma$-critical in a set of GSP's with positive weight in $\kappa_J$, and suppose its coupling value in $J$ is $J_0$. Choose one $\sigma$ within this set, and suppose in this GSP the critical value of $b_0$ is $J_c$; without loss of generality, let $J_0>J_c$.  Because we will be changing (only) the coupling value assigned to $b_0$, we will denote its running  coupling value by $J(b_0)$. As $J(b_0)$ is lowered to $J_c^+$, $\sigma$ will remain unchanged but some fraction (possibly zero) of the GSP's in $\kappa_J$ will have changed due to a critical droplet flip. Focusing only on those GSP's which remain {\it unchanged\/}, Lemmas~2.6 and 3.1 and Theorem~6.3 imply that the process of lowering the coupling value of $b_0$ will have lowered the flexibilities of a positive fraction of edges in a set of $\sigma$'s for which $J(b_0)$ is within the range where $b_0$ controls the flexibilities of a positive fraction of edges. In all of these $P_J(f,\sigma)$ will have changed.

If $\kappa_J$ is supported on a countable (finite or infinite) set of $\sigma$'s, then it is sufficient that $J(b_0)$ is within this range for a single such $\sigma$, given that it will have positive weight in $\kappa_J$.  This will also be true if $\kappa_J$ is supported on an uncountable set of states, if one or more such states also has positive weight in $\kappa_J$. However, a possible scenario is one in which $\kappa_J$ is supported on an uncountable set of states, each of which has {\it zero\/} weight in $\kappa_J$. For this case, the preceding argument needs to be modified.

Suppose then that this scenario holds. Because the edge set $\mathbb{E}^d$ is countable, by the assumption of the theorem a positive density of bonds will be $\sigma$-critical in a set of $\sigma$'s with positive weight in $\kappa_J$. Suppose that the critical values of a bond $b_0$ with this property are dense within an open interval $(J_1,J_2)\subset[-J_{\rm max},J_{\rm max}]$. Choose a coupling value $J_3\in(J_1,J_2)$ and an $\epsilon>0$ and let $J(b_0)\in(J_3,J_3+\epsilon)$. It must be the case that $J(b_0)$ controls the flexibilities of a positive density subset of $\sigma$'s with critical values in this range for {\it some\/} $(J_3,\epsilon)$ pair with $J_3$ and $\epsilon$ chosen as prescribed, by reasoning similar to that used in the last part of the proof in Theorem~6.3.  (This remains true even if there is only a single accumulation point of critical values of $b_0$ in~$(J_1,J_2)$.) Therefore in this scenario $P_J(f,\sigma)$ will also have changed in a set of GSP's with positive weight in $\kappa_J$. 
 
Now assume that $P_J(f,\sigma)$ is independent of $\sigma$.  This would then require two things: first, that in the GSP's in which a critical droplet has {\it not\/} flipped, their $P_J(f,\sigma)$'s have all changed in exactly the same way; and second, that in all $\sigma$'s in which a critical droplet {\it has\/} flipped, as $J_0\to J_c^+$ their final $P_J(f,\sigma)$ has changed identically with those in which no critical droplet flipped. But if these were to happen $P(f)$ will have changed, violating Lemma~7.2. The only remaining alternative is that with $\kappa_J$-measure~one $P_J(f,\sigma)$ cannot be the same for $\kappa_J$-almost every $\sigma$.

\bigskip


\subsection{$\sigma$-criticality and ground state multiplicity}
\label{subsec:finite}

The techniques used above can be taken further to show that the existence of $\sigma$-criticality is incompatible with the existence of metastates supported on a finite number $N>1$ of ground state pairs.



\medskip

{\bf Theorem~7.4.}  Let $N$ denote the number of GSP's on which $\kappa_J$ is supported, and suppose each GSP has a positive fraction of $\sigma$-critical edges.  Then either $N=1$ or $N=\infty$.

\medskip

{\bf Proof.} Assume $2\le N<\infty$. By assumption some bond $b_0$ has positive probability of being $\sigma$-critical in $1\le n\le N$ of the ground state pairs. We can relabel so that this subset of ground state pairs is $\sigma_1, \sigma_2, \ldots \sigma_n$ with $J_{c1}\ge J_{c2}\ge \ldots\ge J_{cn}$, where $J_{ci}$ is the critical value of $J(b_0)$ in ground state pair~$\sigma_i$. 

As in the proof of Lemma~6.4, let $\pm J_{\rm max}$ denote the endpoints of the interval in which $\sigma$-criticality can occur for $b_0$. By Lemma 6.4 and the assumption that there is only a finite number of GSP's in~$\kappa_J$, the intervals $\bigl[J_{c1},J_{\rm max}\bigr]$ and $\bigl[- J_{\rm max},J_{cn}\bigr]$ have nonempty interiors. 
Choose $J_0$ so that $J_{\rm max}>J_0>J_{c1}$.  It follows from Lemmas 2.6 and 3.1 and Theorem~6.3 that lowering $J(b_0)$ from $J_0$ to $J_{c1}^+$ will lower the flexibilities in $\sigma_1$ of a positive density of bonds, and hence will change $P(f,\sigma_1)$. For the remaining $\sigma_2\ldots \sigma_n$, their flexibility distribution will either be changed similarly to that of $\sigma_1$ or else will be unchanged.

As for the other GSP's in the support of $\kappa_J$, $P(f,\sigma_j)$ will remain unchanged, because changing the coupling value of a bond can only change $P(f,\sigma_j)$ if the bond is $\sigma$-critical in $\sigma_j $. Because $n$ is finite, $\sigma_1$ has positive weight in $\kappa_J$, and therefore $P(f)$ will have changed which contradicts~Lemma 7.2. This proves the theorem. 

\bigskip

{\bf Remark.} The proof of Theorem~7.4 can be extended to a countable or uncountable infinity of GSP's if there is a gap in their critical values anywhere inside the interval $\bigl[- J_{\rm max},  J_{\rm max}\bigr]$. In the absence of an argument that such a gap exists, then {\it a priori\/} the set of GSP critical values in either case can be dense in $\bigl[- J_{\rm max},  J_{\rm max}\bigr]$, which would then require a different argument to extend Theorem~7.4 to these cases.

\section{Ground state interfaces and critical droplets}
\label{sec:rsb}

In this section we explore the connection between $\sigma$-criticality and GSP interfaces. Our main result is that a necessary condition for RSB to hold in the EA spin glass at zero temperature is that in every GSP there is a positive probability that any edge is $\sigma$-critical, and a sufficient condition is a positive probability of any edge being $\sigma$-critical of the first kind.

We first make a brief digression. Recall from the discussion in Sect.~\ref{sec:interfaces} that a key feature of the RSB picture is the presence of space-filling, $O(1)$-energy excitations --- i.e., droplets of overturned spins --- above an arbitrary GSP selected from $\kappa_J$. But if the energy of an excitation in $\Lambda_L$ increases no faster than $\sqrt{\vert\partial\Lambda_L\vert}$ {\it and\/} the excitation is space-filling in the infinite-volume limit, the excitation is necessarily an interface between incongruent ground states~\cite{NS02}. We will therefore refer to a space-filling, $O(1)$-energy interface between two infinite-volume GSP's as an {\it RSB interface\/}. It is primarily the existence of RSB interfaces that distinguishes the RSB picture at zero temperature from all others shown in Table~1.

It is important to remember, though, that numerical results~\cite{BY86,PY99b,MP00} indicate that the interface between an {\it arbitrarily selected\/} pair of GSP's chosen from $\kappa_J$ should, in any viable picture, have a high-energy (i.e., increasing with $L$) interface between them. This is not inconsistent with the RSB~prediction that for any GSP $\sigma$, arbitrarily chosen from~$\kappa_J$, there exists a subset of other GSP's that are distant in Hamming space from $\sigma$, but in any finite-volume restriction of $\mathbb{Z}^d$, no matter how large, are separated from it in energy by a gap of $O(1)$.

Returning now to the main discussion, if RSB interfaces exist they should be directly observable. As just noted, numerical evidence~\cite{BY86} strongly indicates that in three and higher dimensions the energy change in switching from periodic to antiperiodic boundary conditions scales as a positive power of the system size, so a different procedure must be used to observe RSB interfaces directly. Fortunately two such methods, which appear to lead to the same results, have been proposed. One is due to Palassini and Young (PY)~\cite{PY00} and the other to Krzakala and Martin~(KM)~\cite{KM00} (see also~\cite{Middleton00}). Although the two methods (to which we refer the reader to the original references) are different, both are designed to observe large-lengthscale excitations with $O(1)$ energy above the ground state. And, not surprisingly, the same outcome was indeed observed in~\cite{PY00} and~\cite{KM00}, namely zero-density excitations of $O(1)$ energy, which led both to propose what is now known as the TNT picture. 

However, these results were disputed in~\cite{MP01}, where use of the PY procedure was claimed to generate positive-density interfaces, i.e., interfaces whose edge set scaled linearly with the volume. We will refer to these finite-volume interfaces between an excitation and ground state as {\it RSB excitations\/}; in the infinite-volume limit they become RSB interfaces between two incongruent GSP's.  If~\cite{MP01} is correct that the PY procedure leads to RSB excitations, then one would expect the KM procedure to do the same. For this reason, when we refer to RSB excitations or interfaces hereafter, we require that they be observable using either the PY or KM methods.

While both methods are expected to lead to the same outcome, for our purposes it is easier to work with the KM procedure, so we briefly describe its essential idea here (adapted to the metastate framework): consider an infinite sequence of volumes $\Lambda_L$ all with periodic boundary conditions. In any given volume, two spins are independently chosen uniformly at random within $\Lambda_L$ and forced to assume a relative orientation opposite to that which they had in the ground state pair $\sigma_L$ (by ``independently'' we mean not only that the two spins are chosen independently of each other in any {\it given\/} volume, but also that the pair of spins is chosen independently {\it between\/} volumes).  The resulting excited state, which we will denote by~$\tau_L$, is then the lowest energy spin configuration in $\Lambda_L$ in which the chosen pair of spins have the opposite orientation from that in $\sigma_L$.

Both the PY and KM methods suggest the following extension of the zero-temperature PBC.  For every volume $\Lambda_L$ collect the following information: $\sigma_L$, $\tau_L$ (which will refer to the excited state in either procedure), $\Delta E_L = E_L(\tau_L)-E_L(\sigma_L)>0$, and the edge set corresponding to the interface between $\sigma_L$ and $\tau_L$. Then because the joint distribution~$(J_L,\sigma_L,\tau_L)$ is invariant under torus translations of $\Lambda_L$, the limiting distribution~$\mu_J$ is translation-invariant on~$\mathbb{Z}^d$ and is itself a kind of metastate containing more information than $\kappa_J$ (in fact, $\mu_J$ is a special case of the excitation metastate).

What is the significance of the $\tau_L$? It turns out that {\it in the infinite-volume limit, these all become ground states themselves\/}. It is not hard to see why this is so. Working within the KM procedure, fix a finite volume~$\Lambda_{L_0}$, which will serve as a ``window'' in the sense described in Sect.~\ref{subsec:pbc}, and study the excited spin configurations inside it generated within~$\Lambda_L$ (with $L\gg L_0$ always). The independently-chosen spins will then move outside of $\Lambda_{L_0}$ with probability approaching~one as $L\to\infty$. 

Consider one of these $\Lambda_L$, and call the two independently chosen spins $\sigma_1$ and $\sigma_2$; $\tau_L$ is then the lowest-energy configuration in $\Lambda_L$ subject to $\sigma_1$ and $\sigma_2$ having the opposite relative orientation to what they had in $\sigma_L$. But then Eq.~(\ref{eq:gs}) must hold for any contour or surface completely inside $\Lambda_L$ that includes either {\it both\/} or {\it neither\/} of $\sigma_1$ and $\sigma_2$. Because $\sigma_1$ and $\sigma_2$ eventually move outside any fixed $L_0$ in the infinite-volume limit, Eq.~(\ref{eq:gs}) becomes satisfied in $\tau_L$ for every closed contour or surface inside {\it any\/} window of fixed size, no matter how large. Therefore any infinite-volume spin configuration $\tau$ which is a convergent subsequence of $\tau_L$'s satisfies the definition of an infinite-volume GSP.

Although the argument above was done using the KM procedure, a similar argument leads to the same result for PY. While we exclusively use the KM procedure in this paper, all results obtained for KM should hold equally well for PY.  So if, as argued in~\cite{MP01}, these procedures generate space-filling interfaces, then $\mu_J$ will be supported on {\it pairs\/} of GSP's separated by RSB interfaces.

An interesting question arises: although these $\tau$'s are in the support of $\mu_J$, are they also in the support of the original $\kappa_J$? In other words, we know from the above argument that the $\tau$'s are GSP's of the infinite-volume EA Hamiltonian~(\ref{eq:EA}). But are they also subsequence limits of an infinite sequence of finite-volume GSP's with periodic boundary conditions? Of course, if all coupling-independent and spin-flip-symmetric boundary condition metastates are the same --- which is strongly suspected~\cite{NS98} but not yet rigorously proved --- then the question is immediately answered in the affirmative. In the absence of a proof though, it would be useful to know whether the $\tau$'s have this property. The next theorem shows that they do.

\medskip

{\bf Theorem 8.1.} If the limiting infinite-volume spin configurations $\tau$ generated from the excited states~$\tau_L$ using the KM or PY procedure have RSB interfaces with the corresponding $\sigma$'s, then they are infinite-volume ground states also in the support of $\kappa_{J}$.

\medskip

{\bf Proof.} The argument above already showed, using the condition~(\ref{eq:gs}), that the $\tau$'s are ground states of the infinite-volume Hamiltonian~(\ref{eq:EA}). What remains to be shown is that the $\tau$'s are also in the support of~$\kappa_{J}$. Suppose this is not the case. One can then construct a new metastate $\kappa'_{J}\ne\kappa_{J}$ supported solely on the set of $\tau$'s. By the assumption of the theorem, these $\tau$'s are incongruent with $\sigma$'s in the support of $\kappa_{J}$. Now select from $\mu_J$ a $(\sigma,\tau)$ pair separated by an RSB interface, with $\sigma$ in the support of $\kappa_{J}$ and $\tau$ in the support of $\kappa'_{J}$, and examine the fluctuations in their energy difference $\Delta E_{L}$ within any volume~$\Lambda_L\subset\mathbb{Z}^d$. Using Theorem~3.3 from \cite{ANSW16}, there is a constant $c>0$ such that for any~$\Lambda_L$ sufficiently large, Var$(\Delta E_{L})\ge c\vert\Lambda_L\vert$, which violates the condition that the energy difference between the selected $\sigma$ and $\tau$ is never greater than $O(1)$ in any $\Lambda_L$.

\bigskip

{\bf Remark.} The proof of Theorem 8.1 shows that the conclusion remains valid for a large class of procedures that generate infinite-volume $\tau$'s that are incongruent with some set of $\sigma$'s in the support of $\kappa_J$.

\medskip

We turn now to the relation between $\sigma$-criticality and RSB interfaces. We begin by showing that a sufficient condition for the presence of~RSB interfaces between a GSP~$\sigma$ and (one or more) other GSP's is that a positive density of bonds have space-filling critical droplets in~$\sigma$. 

\medskip

{\bf Theorem 8.2.} If a GSP $\sigma$ chosen from $\kappa_J$ has a positive fraction of edges with space-filling critical droplets, then $\sigma$ will have an RSB interface with one or more other GSP's in~$\kappa_J$.

\medskip

{\bf Proof.}  We introduce the following procedure for generating excited states in a given volume $\Lambda_L$. Choose an arbitrary~{\it bond\/} uniformly at random within~${\mathbb E}_L$ (the edge set restricted to $\Lambda_L$) and consider the excited state~$\tau_L$ generated by flipping its critical droplet (with $J$ remaining fixed).  As before, the bond is chosen independently for each~$\Lambda_L$.

By assumption, in any $\Lambda_L$ the procedure defined above has a positive probability of generating a positive-density critical droplet, so there is a set of $\tau_L$ with positive measure in~$\kappa_J$ generated by this procedure having a positive-density interface with the corresponding~$\sigma_L$.  By the usual compactness arguments the set of interfaces between the $\tau_L$'s and $\sigma_L$'s will converge to limiting space-filling interfaces in one or more subsequences of $\Lambda_L$'s.  By construction the energy of the interface in any volume is twice the flexibility of the chosen bond, so in the infinite-volume limit the energy of the generated interface between $\tau$ and $\sigma$ remains~$O(1)$ in any finite-volume subset of ${\mathbb Z}^d$.

Consider one such bond~$b_1$ chosen in $\Lambda_L$ which by this procedure generates a $\tau_L$ having a positive-density interface with $\sigma_L$.  By definition the critical droplet is the lowest-energy droplet generated by changing a bond's coupling value past its critical value.  Then the condition~(\ref{eq:gs}) is satisfied in the $\tau_L$ generated by flipping the critical droplet of $b_1$ for all closed contours or surfaces {\it except\/} those passing through $b_1$.  But by the same arguments as those leading to Theorem~8.1, the chosen bond will move outside any fixed window with probability approaching one as $L\to\infty$; so in any fixed volume~(\ref{eq:gs}) will be satisfied in $\tau_L$ for sufficiently large $L$.  Consequently any infinite-volume~$\tau$ generated by this procedure is itself an infinite-volume GSP of the Hamiltonian~(\ref{eq:EA}), and by the remark following the proof of Theorem~8.1 it is in the support of~$\kappa_J$.

\bigskip

Theorem 8.2 shows that the presence of space-filling critical droplets (which by Theorem~6.2 are $\sigma$-critical) is a sufficient condition for RSB interfaces between GSP's to be present within $\kappa_J$. We now consider a necessary condition. 

\medskip

{\bf Theorem 8.3.} If a GSP~$\sigma$ chosen from $\kappa_J$ has an RSB interface with one or more other GSP's in $\kappa_J$, then a positive fraction of edges in $\sigma$ are $\sigma$-critical.

\medskip

{\bf Proof.}  If RSB holds in a given dimension, then by prediction RSB excitations will be observable using the PY or KM procedure in a positive fraction of volumes~\cite{MP01}; here we will use the KM procedure. Let $\Lambda_L$ be one of the volumes in which an RSB excitation is generated and call the two spins chosen by the KM procedure $\sigma_1$ and $\sigma_2$. As before call the original GSP~$\sigma_L$ and the excited state pair~$\tau_L$, and let ${S}_{\rm KM}$ denote the interface between $\sigma_L$ and $\tau_L$; $S_{\rm KM}$ is an RSB excitation.

Next choose a bond $b_2\in S_{\rm KM}$ and consider its critical droplet $D(b_2,\sigma_L)$. If the droplet corresponding to $D(b_2,\sigma_L)$ includes only one of $(\sigma_1,\sigma_2)$ and its boundary $\partial D(b_2,\sigma_L)$ is different from ${S}_{\rm KM}$ then there's a contradiction, because by definition $E(\partial D(b_2,\sigma_L))<E_{{S}_{\rm KM}}$, where both energies are computed in $\sigma_L$ as in~(\ref{eq:gs}). Therefore, if the critical droplet of~$b_2$ includes only one of $(\sigma_1,\sigma_2)$, it is already space-filling.  As a consequence, in what follows we will assume that $\partial D(b_2,\sigma_L)$ encloses both or neither of $\sigma_1$ and $\sigma_2$.

Let $\tilde{{S}}={S}_{\rm KM}\cap\partial D(b_2,\sigma_L)$; $\tilde{{S}}$ contains at least $b_2$ but could contain other bonds as well. Split $\partial D(b_2,\sigma_L)$ into two pieces, $\tilde{{S}}$ and $\partial D^<$, such that $\tilde{{S}}\cap\partial D^<=\emptyset$ and $\tilde{{S}}\cup\partial D^<=\partial D(b_2,\sigma_L)$. Do the same with ${S}_{\rm KM}$, so that $\tilde{{S}}\cap S_{\rm KM}^<=\emptyset$ and $\tilde{{S}}\cup S_{\rm KM}^<=S_{\rm KM}$.

Because $b_2$ belongs to both ${S}_{\rm KM}$ and $\partial D(b_2,\sigma_L)$, $E(\partial D(b_2,\sigma_L))<E({S}_{\rm KM})$ so 
\begin{equation}
\label{eq:energies1}
E\bigl(\partial D(b_2,\sigma_L)\bigr)=E({\tilde S})+E(\partial D^<)<E(S_{\rm KM})=E(\tilde{S})+E({S}_{\rm KM}^<)\Rightarrow E(\partial D^<)<E({S}_{\rm KM}^<)\, .
\end{equation}

Additional information can be gained by noting that any surface enclosing a droplet which includes either both or neither of $\sigma_1$ and $\sigma_2$ must have positive energy when computed in either $\sigma_L$ or $\tau_L$; it follows that $E(\partial D(b_2,\sigma_L))>0$ in both the ground and KM spin configurations.  Because $E(\tilde{S})$ has the same magnitude and opposite sign in the two states, it follows that $E(\partial D^<)>0$. We therefore have the inequalities
\begin{equation}
\label{eq:energies2}
0<E(\partial D^<)<E({S}_{\rm KM}^<)\, .
\end{equation}

Next consider another bond $b_3$ in ${S}_{\rm KM}^<$ (and therefore not in~$\partial D^<$). Its coupling magnitude can be changed (in a direction to bring it closer to its critical value in $\sigma_L$), and the energy $E({S}_{KM}^<)$ correspondingly lowered, by any amount up to its flexibility $f(b_3,\sigma_L)$ without affecting either the ground or KM states. Such an operation may or may not lower $E({S}_{\rm KM}^<)$ below $E(\partial D^<)$. Suppose it does not. How often can we repeat this procedure without creating a droplet flip? 

Suppose a zero fraction of edges are $\sigma$-critical; as a consequence if $S_{KM}$ is space-filling, then any bond whose dual is in $\partial S_{\rm KM}$ has a critical droplet different from $S_{\rm KM}$. Let $P_{\sigma_L}(b_{i_1},b_{i_2},\ldots,b_{i_n})$, with $n$ fixed and finite, be the probability in $\Lambda_L$ that any bond in the set~$\{b_{i_1},b_{i_2},\ldots,b_{i_n}\}$ is in the critical droplet boundary of at least one other bond in the set. If once again we choose the $n$ bonds randomly and independently in each volume $\Lambda_L$, then the absence of $\sigma$-criticality would ensure that $P_{\sigma_L}(b_{i_1},b_{i_2},\ldots,b_{i_n})\to 0$ as $L\to\infty$, for any fixed, finite $n$.  

This conclusion is unchanged if we restrict the chosen bonds to belong to $S^<_{\rm KM}$ in any volume in which $S_{\rm KM}$ is space-filling: because of the absence of $\sigma$-criticality, $\partial D(b_2,\sigma_L)$ cannot be space-filling, so ${\tilde S}$ cannot be either; therefore $S^<_{\rm KM}$ must be.


Using the reasoning above, the number $n$ of such ``non-interacting'' sets of bonds in $S_{\rm KM}$ can slowly increase to infinity as $\Lambda_L\to\infty$. Changing the magnitude of the coupling associated with one of these bonds has no effect on the flexibility of any of the others as long as no droplet flip occurs in~$\sigma_L$, so the operation described above can be performed on an increasing number of bonds (as $L$ increases) in $S_{\rm KM}$. 

Therefore, because $E({S}_{\rm KM}^<)-E(\partial D^<)$ was $O(1)$ before changing any coupling values, and the flexibility of every individual bond is also $O(1)$, and we can perform the operation described above on $b_3$ independently on a number $n$ of ``non-interacting'' bonds (with the original $b_2$ included in this set) that increases with the size of the volume under consideration, it follows that the inequality~(\ref{eq:energies1}) must be eventually violated in sufficiently large volumes.  As a consequence, the presence of RSB interfaces is incompatible with the absence of $\sigma$-criticality.

\bigskip



Theorems~8.2 and 8.3 can be combined as the following theorem:

\medskip

{\bf Theorem 8.4.} RSB interfaces between ground states in the zero-temperature periodic boundary condition metastate~$\kappa_J$ are present if each $\{\sigma$\} has a positive density of bonds which have a space-filling critical droplet, and only if each $\sigma$ has a positive density of bonds which are $\sigma$-critical.

\medskip

We conclude our discussion on the connection between RSB and $\sigma$-criticality with a theorem on ground state multiplicity in the RSB picture.  It is well known that RSB is a many-state picture~\cite{MPV87} based on the correspondence between replicas and pure states~\cite{Parisi83,MPV87}, and it has been rigorously established that the positive temperature PBC~metastate is supported on an uncountable infinity of pure states~\cite{NS06b}; but we are unaware of any rigorous results on GSP multiplicity in the RSB picture at zero temperature. (There is, however, a rather surprising result~\cite{NS03c} about GSP multiplicity in the infinite-range SK~model~\cite{SK75}, in which it was shown that with appropriate subsequence limits {\it every\/} infinite-volume spin configuration is a ground state.) With the correspondence between RSB and $\sigma$-criticality established in this section, however, we can now state the following theorem:

\medskip

{\bf Theorem 8.5.} If RSB holds in some dimension, then in that dimension the zero-temperature PBC metastate $\kappa_J$ is supported on infinitely many incongruent GSP's.

\medskip

{\bf Proof.} By the Theorem in Sect.~6 of~\cite{NS02} and Theorem~1 of~\cite{AFunpub,NSunpub,Stein16}, the presence of space-filling interfaces with energy scaling no faster than $L^{(d-1)/2}$ along the interface requires $\kappa_J$ to be supported on $N\ge 2$ incongruent ground state pairs. By Theorem~8.3, if RSB holds in some dimension, then there is positive probability that any edge is $\sigma$-critical. Theorem 7.4 then requires that either $N=1$ or $N=\infty$; therefore $N=\infty$.

\bigskip

Theorem 8.5 does not specify whether $\kappa_J$ is supported on a countable or uncountable infinity of GSP's. Based on the rigorous result of an uncountable infinity of pure states at positive temperature~\cite{NS06b}, however, we are prepared to make the following conjecture: 

\medskip

{\bf Conjecture 8.6.} If RSB holds in some dimension, then in that dimension the zero-temperature PBC metastate $\kappa_J$ is supported on an uncountable infinity of incongruent GSP's.

\section{Critical droplets and ground state multiplicity}
\label{sec:multi}

We saw in the previous section that $\sigma$-criticality of the first kind --- i.e., a positive probability of an edge having a space-filling critical droplet in a ground state --- is a sufficient condition for RSB to hold in the EA model in a given dimension. While the presence of $\sigma$-criticality of the second kind would be enough to provide a necessary condition, it is unclear whether it is also sufficient for RSB to be present. One might then speculate whether the presence of $\sigma$-criticality of the second kind combined with absence of the first kind might lead to a different picture, such as chaotic pairs, or even (less plausibly) scaling-droplet or TNT.

To address this question we take a step back and ask whether necessary or sufficient conditions can be found for multiple incongruent GSP's to be present at all, {\it regardless\/} of which picture results.  

As in earlier sections, our interest is in GSP multiplicity that is observable using straightforward and standard physical procedures. The usual approach is to ask whether changing the boundary conditions ``at infinity'' can change the spin configuration (other than a global flip) inside a large but fixed finite window centered at the origin. Here ``changing the spin configuration'' 
means that some bond inside the window changes its state from satisfied to unsatisfied or vice-versa.\footnote{This means that with positive probability the number of edges undergoing such a change scales like the volume of the window.}

This leads to the following definition:

\medskip

{\bf Definition 9.1.} Consider an infinite sequence of volumes $\Lambda_L$, with $L\to\infty$. If, in a positive fraction of these volumes, a change in boundary condition from periodic to antiperiodic changes the spin configuration inside any (large, but small compared to $\Lambda_L$) fixed window centered at the origin, then we will say that the zero-temperature periodic boundary condition metastate~$\kappa_J$ is supported on observably multiple ground state pairs. 

\medskip

{\bf Remark}. As noted in Sect.~\ref{sec:chaos}, it was proved in~\cite{NS01c} that the distinct ground state pairs generated in this way are necessarily mutually incongruent. In the rest of the paper  ``multiple ground state pairs'' will mean observably multiple.

\medskip

It will be important to keep in mind that with probability~one any two metastates generated by choosing {\it either\/} periodic or antiperiodic boundary conditions for each $\Lambda_L$, independently of the couplings and $L$, are equal to each other~\cite{NS98}.  The method of proof further implies a stronger result, namely that the same is true for boundary conditions that, in any {\it single\/} volume $\Lambda_L$, choose (again independently of the couplings, etc.) periodic boundary conditions for some pairs of opposite boundary spins and antiperiodic for others. By ``opposite boundary spins'' we mean spins $(\sigma_1,\sigma_2)$ that have the same coordinates in $(d-1)$~dimensions and coordinates ($-L/2,+L/2)$ in the $d^{\rm th}$ dimension. This stronger result will be useful in what follows, and will be included as a theorem below after some preliminary discussion.

\medskip

We first take a closer look at the process of switching from periodic to antiperiodic boundary conditions along a {\it single\/} direction in a specific volume. Consider the edge set $E_L$ comprising all bonds connecting two sites $x$ and $y$ with $x$ and $y$ both in~$\Lambda_L$, and the set $\partial E_L$ consisting of bonds connecting sites $u$ and $v$ with $u\in\Lambda_L$ and $v\in\partial\Lambda_L$. The two sets are distinct; we will refer to bonds belonging to~$E_L$ as ``interior bonds'' and those belonging to $\partial E_L$ as ``boundary bonds''.

It has long been recognized that switching from periodic to antiperiodic boundary conditions along a single direction is equivalent to restoring periodic boundary conditions everywhere on $\partial\Lambda_L$ while simultaneously reversing the signs of all couplings $J_{xy}$ corresponding to boundary bonds in $\partial E'_L$, where $\partial E'_L\subset\partial E_L$ consists of the set of boundary bonds along a single side (in two dimensions) or face (in higher dimensions) belonging to $\partial\Lambda_L$.  This result follows from the invariance of~(\ref{eq:EA}) with respect to the gauge transformation $\sigma_x\to-\sigma_x$ simultaneously with $J_{xy}\to -J_{xy}$ for all $y$ satisfying $\| x-y\| =1$.

An immediate consequence of the above observation is that the change of spin configuration inside $\Lambda_L$, due to switching the signs of the $J_{xy}$ corresponding to boundary bonds $b_{xy}$ in $\partial E'_L$, is determined by the critical droplet size distribution in ground states belonging to the support of $\kappa_J$. To see this, note that a global change in sign of all couplings along a face of $\partial\Lambda_L$ yields the same interior configuration as locally changing the same set of couplings one at a time, each time letting the spins rearrange to the new GSP, and going on to the next. After every coupling change from $J_{xy}\to -J_{xy}$ in a boundary bond, one of two things happens: either the bond's critical value is not crossed, so the bond's critical droplet does not flip, and the GSP remains unchanged; or else\footnote{Some coupling's sign reversal will lead to a critical droplet flip in any GSP; otherwise~(\ref{eq:gs}) will be violated if all couplings along a closed surface reverse sign.} the bond's critical value is crossed, flipping the bond's critical droplet thereby leading to a new GSP inside $\Lambda_L$ (but not necessarily changed inside the window).  As already noted, the GSP following each coupling sign reversal is the same as would result with the original~$J$ and a change of boundary condition from periodic to antiperiodic on the corresponding opposite spin pair. Sign reversals notwithstanding, any new GSP generated in this way remains in the support of the original $\kappa_J$, as a consequence of the following theorem:

\medskip

{\bf Theorem 9.2 (Newman-Stein)~\cite{NS98}.} Any two metastates generated by an infinite sequence of volumes with different boundary conditions are the same if the two boundary conditions on each volume are related by a gauge transformation.

\medskip

Clearly, if changing from periodic to antiperiodic BC's on only a subset of opposite spin pairs in a positive fraction of volumes leads to a change of spin configuration inside a fixed window, then there also exist multiple incongruent GSP's in the support of $\kappa_J$. This leads to the following theorem:

\medskip

{\bf Theorem 9.3.} The presence of $\sigma$-criticality of {\it either\/} kind is a sufficient condition for the existence of multiple observable ground state pairs in the support of~${\kappa}_J$.

\medskip

{\bf Proof.} Consider a volume $\Lambda_L$ along with a fixed window $\Lambda_{L_0}$, both centered at the origin and with $1\ll L_0\ll L$.  By Definition~9.1 and the remark following, if there are two or more incongruent GSP's in the support of $\kappa_J$ then there is positive probability that a change from periodic to antiperiodic boundary conditions for one or more pairs of opposite boundary spins along one direction will effect a change in the spin configuration inside $\Lambda_{L_0}$.  

From Theorem~8.2 we already know that $\sigma$-criticality of the first kind is a sufficient condition for GSP multiplicity.  Suppose this is absent but $\sigma$-criticality of the second kind is present. From the discussion above, switching any pair of opposite boundary spins from periodic to antiperiodic is equivalent to changing the sign of the coupling of a single bond in $\partial E'_L$. Let $E_{L_0}$ denote the set of all bonds $b_{xy}$ such that $x\in\Lambda_{L_0}$ and $y\in\Lambda_{L_0}$. Because of the presence of $\sigma$-criticality of the second kind, a bond has positive probability of belonging to the critical droplet boundary of a positive density of bonds in~$\mathbb{E}_L$; so changing the sign of a coupling on a boundary bond arbitrarily far away has positive probability of causing a bond in $E_{L_0}$ to change its state from satisfied to unsatisfied or vice-versa. There is then positive probability that the spin configuration inside $\Lambda_{L_0}$ will change when the BC's of any $\Lambda_L$ (with $L>L_0$) are changed from periodic to (partially or fully) antiperiodic. This completes the proof.

\bigskip

We next consider a necessary condition for observing incongruent GSP's in $\kappa_J$.  As discussed above, the equivalence between changing boundary conditions and flipping the signs of boundary couplings one at a time suggests that the observation of multiple GSP's requires the existence of critical droplets of arbitrarily large size.  This will certainly occur if $\sigma$-criticality is present, but it might also occur in its absence. 

To proceed, we introduce the following quantities. Let $K^*(b,\sigma)$ denote the number of bonds in ${\mathbb E}^d$ whose critical droplet boundaries in $\sigma$ pass through $b$. Then
for $k = 1,2,3\ldots$ define $P(k, \sigma)$ to be the fraction of bonds~$b\in{\mathbb E}^d$ such that $K^*(b,\sigma) = k$, and let 
\begin{equation}
\label{eq:avg}
E_\sigma [K^*] = \sum_{k=1}^\infty k\ P(k, \sigma)\, .  
\end{equation}
That is, $E_\sigma[K^*]$ is the average number of bonds whose critical droplet boundaries a typical bond belongs to in the GSP $\sigma$. We then define the following:

\medskip

{\bf Definition 9.4.} A ground state $\sigma$ will be called $\sigma$-{\it subcritical\/} if $\sigma$-criticality of either kind is absent, but $E_\sigma[K^*]=\infty$.

\medskip

{\bf Remark.}  A GSP will be $\sigma$-subcritical if, for instance, a positive fraction of bonds belong to the critical droplet boundaries of an infinite, {\it zero-density\/} set of other bonds. Neither $\sigma$-criticality nor $\sigma$-subcriticality will be present if $P(k,\sigma)$ falls off faster than $k^{-(2+\epsilon)}$ for any $\epsilon>0$ as $k\to\infty$.

\medskip

If $\sigma$-criticality and $\sigma$-subcriticality are both absent, i.e., $E_\sigma[K^*]<\infty$, then a fraction one of all bonds belong to the critical droplet boundary of only a finite number of other bonds (the converse of course is not necessarily true). If this were to occur, then, as we argue below, the spin configuration in any fixed finite window should not be affected by changing the boundary condition from periodic to antiperiodic in a volume whose boundaries are sufficiently far from the window. 

To test this conjecture, we define the following. As in the proof of Theorem~9.3, consider a volume $\Lambda_L$ along with a fixed window $\Lambda_{L_0}$, both centered at the origin and with $1\ll L_0\ll L$.  The underlying probability space for the discussion here is the one corresponding to the choice of the coupling configuration $J$.  Then let ${\cal A}_L$ denote the event that switching from periodic to antiperiodic boundary conditions, either along all of $\partial E'_L$ or else a subset of~$\partial E'_L$, changes the spin configuration inside the window $\Lambda_{L_0}$ such that there is some subset of bonds belonging to $E_{L_0}$ which change their status from satisfied to unsatisfied, or vice-versa, in the new ground state spin configuration.

Next, order the $n$ couplings in $\partial E'_L$ from 1 to $n$ (which index is assigned to which particular coupling in $\partial E'_L$ is immaterial so long as every coupling is assigned one and only one index, and each coupling's index is distinct from all the others). Let ${\cal B}_{L,i}$, $i=1,\ldots, n$ be the event that, with periodic boundary conditions on $\partial\Lambda_L$, changing the sign of the {\it single\/} coupling $J_i$ assigned to a bond $b_i\in\partial E'_L$, while leaving all other couplings unchanged from $J$, changes the spin configuration inside $\Lambda_{L_0}$ such that there is some subset of bonds belonging to $E_{L_0}$ which change their status from satisfied to unsatisfied, or vice-versa, in the new ground state spin configuration. In other words, the event ${\cal B}_{L,i}$ occurs when the critical droplet of~$b_i\in\partial E'_L$ penetrates the window $\Lambda_{L_0}$.

Finally, again with periodic boundary conditions on $\partial\Lambda_L$, reverse the sign of couplings $1,\ldots,i$, one at a time in order of index. The event ${\cal C}_{L,i}$ then occurs if changing the sign of the $i^{\rm th}$ boundary coupling changes the spin configuration inside $\Lambda_{L_0}$ from its configuration just before the $i^{\rm th}$ coupling sign was flipped.

Several relations follow from the above definitions, for a fixed volume $\Lambda_L$ with $n=\vert\partial E'_L\vert$ boundary bonds along one face:

\medskip

(a) ${\cal A}_L=\bigcup_{i=1}^{n}{\cal C}_{L,i}$;

\smallskip

(b) ${\cal B}_{L,1}={\cal C}_{L,1}$;

\smallskip


\smallskip

(c) Using gauge-invariance as above, $P({\cal B}_{L,i})=P({\cal C}_{L,i})$.

\smallskip

From (a) and (c) above we obtain the union bound 
\begin{equation}
\label{eq:unionbound}
P({\cal A}_L)=P\Bigl(\bigcup_{i=1}^{n}{\cal C}_{L,i}\Bigr)\le\sum_{i=1}^{n}P({\cal C}_{L,i})=\sum_{i=1}^{n}P({\cal B}_{L,i})\, .
\end{equation}

Now suppose that $\kappa_J$ is supported on multiple incongruent GSP's.  If so, then switching from periodic to antiperiodic boundary conditions changes the spin configuration inside $\Lambda_{L_0}$ in a positive fraction of $\Lambda_L$'s.  Thus for a large $\Lambda_L$, $P({\cal A}_L)>c>0$, where $c$ is independent of $L$, and consequently 
\begin{equation}
\label{eq:bound}
\sum_{i=1}^{n}P({\cal B}_{L,i}) > c>0
\end{equation}
independently of $L$.



Now let $b_0$ be a fixed bond and $b'$ be a varying bond located anywhere in $\mathbb{E}^d$.
Let ${\cal D}_{b',b_0}$ denote the event that the boundary of
the critical droplet in~$\sigma_J$ caused by modifying the value of $J_{b'}$ passes through $b_0$, 
and let $P({\cal D}_{b',b_0})$ denote the probability of occurrence (with respect to $(J,\sigma_J)$) of ${\cal D}_{b',b_0}$.

Then (\ref{eq:bound}) suggests (see the remark following Claim~9.5 below) that
\begin{equation}
\label{eq:bound2}
P({\cal D}_{b',b_0}) > C\Vert b'-b_0\Vert^{-(d-1)}
\end{equation}
where $0<C<\infty$ is some constant and $\Vert b'-b_0\Vert$ denotes the Euclidean distance between $b'$ and $b_0$.

Finally, if $E_{J,\sigma}[\cdot]$ denotes the average both over couplings $J$ and $\sigma$'s from $\kappa_J$, then~(\ref{eq:bound2})
implies that $E_{J,\sigma}[K^*(b_0,\sigma)] =\infty$, which in turn suggests that $E_\sigma [K^*] = \infty$ for (at least) a positive fraction of the $\sigma$'s in 
the metastate for (at least) a positive fraction of the $J$'s.  We therefore conclude with the following claim:

\medskip

{\bf Claim 9.5.} A necessary condition for $\kappa_J$ to be supported on multiple GSP's is that $E_\sigma[K^*]=\infty$ --- i.e., either $\sigma$-criticality or $\sigma$-subcriticality is present ---- in a positive fraction of the GSP's in the support of $\kappa_J$.

\medskip

{\bf Remark.} Although the informal argument above provides convincing evidence that $E_\sigma[K^*]=\infty$ is a necessary condition for multiple GSP's to be in the support of $\kappa_J$, we present it as a (nonrigorous) claim rather than a rigorous theorem. This is because the quantities related to $K^*$ are defined with respect to infinite-volume ground states in the metastate, while~(\ref{eq:bound}) is a condition on finite-volume ground states. (This is similar to the issue discussed following Eqs.~(\ref{eq:flucs1} )and~(\ref{eq:oldflucs}) in Sect.~\ref{sec:chaos}.) One could of course define finite-volume equivalents of $K^*$ and related quantities, but the point of view of this paper is that a coherent picture requires all results to be formulated within the metastate framework.

\section{Summary and Discussion}
\label{sec:discussion}

The nature of the spin glass phase has not been settled despite years of investigation and a vast literature. Numerous pictures have been proposed, which differ on fundamental aspects of spin glass equilibrium behavior: multiplicity of pure states at low temperature or ground states at zero temperature; the number and structure of positive-temperature thermodynamic states (i.e., whether each one in the support of the metastate comprises a single pure state pair or a nontrivial mixture of infinitely many incongruent pure state pairs); interface structure and energy; the geometry and energetics of the lowest-energy large-lengthscale excitations above the ground state; and so on. 

In this paper we proposed that all of these features are related to the stability of the ground state to a change in coupling value of a {\it single\/} bond. This is different from but, as shown in~\cite{ANS19} and discussed in~Sect.~\ref{sec:chaos}, can be related to the concept of disorder chaos~\cite{BM87,FH88b,KB05,KK07,Chatterjee09,ANS19}. In this paper the primary objects of interest are the flexibility of a coupling, which is the amount by which the coupling can be varied before forcing a change in the ground state of interest, and the associated bond's critical droplet, which is the droplet flip caused by changing the bond's coupling value. Flexibilities, critical droplets, and associated quantities can be defined both for finite-volume and infinite-volume ground states. There are three spin glass systems, discussed in Sect.~\ref{subsec:special}, for which these quantities are completely or partially understood: the EA~model in one dimension, the highly disordered model in all dimensions, and the strongly disordered model in all dimensions.

Our main interest is the EA spin glass within the context of the zero-temperature periodic boundary condition metastate~$\kappa_J$, defined in Sect.~\ref{subsec:pbc}. We focused on four prominent conjectures about the nature of the spin glass phase in this model: RSB, scaling-droplet, chaotic pairs, and TNT.  These can be categorized and related by the geometry and energy of the large-lengthscale, low-energy excitations above the ground state at zero temperature (see Table~1 in Sect.~\ref{sec:interfaces}).  In order to investigate the connection between these pictures and ground state stability, we introduced the concept of $\sigma$-criticality, which measures the extent to which changing the coupling value of a single bond affects the flexibilities of other bonds in the infinite-volume edge set~$\mathbb{E}^d$. If a bond is $\sigma$-critical, then lowering its flexibility by varying its coupling (within some open interval) lowers the flexibilities of a positive density (in $\mathbb{E}^d$) of other bonds.

This can arise in two ways: $\sigma$-criticality of the first kind occurs when the critical droplet of a bond is space-filling (i.e., has a positive-density boundary); $\sigma$-criticality of the second kind occurs when a bond's critical droplet is not space-filling but it belongs to the critical droplet boundary of a positive density of bonds in $\mathbb{E}^d$. One way in which the second kind could occur is if a positive density of bonds have infinite but zero-density critical droplets. 

A bond can also be $\sigma$-subcritical, meaning that lowering its flexibility also lowers the flexibility of an infinite but {\it zero-density\/} set of other bonds. One of our main results, presented in Sect.~\ref{sec:multi}, is that a sufficient condition for ground state pair multiplicity to arise is that GSP's in the support of $\kappa_J$ have a positive density of edges being $\sigma$-critical (of either kind), and a likely necessary condition is that GSP's have a positive density of edges being either $\sigma$-critical or $\sigma$-subcritical. 

This last result is a general statement for the EA model and is independent of which picture one assumes describes the spin glass phase. We now turn to the four pictures summarized in Table~1 and how each relates to a different type of ground state stability.  It is important to emphasize that Table~1 refers solely to the large-lengthscale excitations which dominate the zero- and low-temperature behavior of the spin glass phase.  We already noted in Sect.~\ref{sec:rsb} that {\it high-energy\/} space-filling interfaces must also be present in the RSB picture. In principle such interfaces could also coexist with either TNT-type excitations (i.e., infinite, zero-density low-energy excitations) or scaling-droplet-type high-energy compact excitations.  But if this were to happen in either case it would then be an example of the chaotic pairs picture, in which $\kappa_J$ is supported on multiple GSP's and the low-temperature PBC metastate is supported on many thermodynamic states, each of which comprises a single pure state pair.

To avoid confusion (and adhere to the original definitions of these pictures), when we refer to either the TNT or scaling-droplet pictures we mean that the excitations they predict are the {\it only\/} ones that determine the ground state and low-temperature properties of the spin glass phase. Specifically, this means that {\it no\/} space-filling excitations of either energy type~(with energy scaling no faster than $O(L^{(d-1)/2})$) are present in either the TNT or scaling-droplet pictures. We further require, as discussed in the text, that all excitations must be observable, through either the PY/KM procedures in the case of RSB and TNT, or through a change of periodic to antiperiodic boundary conditions in the chaotic pairs or scaling-droplet pictures.

With these clarifications in mind, we now discuss the relation of each of the four pictures in Table~1 to ground state stability, beginning at the upper left. 
In Sect.~\ref{sec:rsb} we showed that a necessary condition for the presence of RSB is that $\sigma$-criticality of either kind be present in GSP's in the support of $\kappa_J$, while a sufficient condition is the presence of $\sigma$-criticality of the first kind. Because of this sufficient condition, none of the other pictures is compatible with the presence of $\sigma$-criticality of the first kind.

The chaotic pairs picture, then, must be a consequence of the presence of either $\sigma$-criticality of the second kind or $\sigma$-subcriticality.

For both the TNT and scaling-droplet pictures, $\sigma$-criticality of the second kind is also ruled out, since its presence implies multiplicity of incongruent GSP's. At this time we cannot rule out the incompatibility of $\sigma$-subcriticality with either picture --- though the argument supporting Claim~9.5 suggests the possibility that $\sigma$-subcriticality might also be a sufficient condition for the existence of GSP multiplicity in $\kappa_J$, in which case its presence too would be incompatible with either the TNT or scaling-droplet pictures.

The TNT picture is compatible with at least two possibilities. If TNT excitations really are infinite, they could in principle be generated from an infinite but zero-density set of bonds having infinite, zero-density critical droplets. But it could also be the case --- consistent with the numerical data in~\cite{PY00} and~\cite{KM00} --- that all critical droplets are {\it finite\/} but with a size distribution falling off slowly with increasing length.  




As for the scaling-droplet picture, it requires finite critical droplets only.  Its scaling predictions at very low energies~\cite{FH86,FH88b} imply a power-law falloff in the linear extent of compact droplet excitations with $O(1)$ energy; this then implies a similar falloff in size for the critical droplet distribution.


An interesting possibility, discussed earlier in (Sect.~\ref{sec:chaos}), is the case of an exponential falloff in the distribution of sizes at large lengthscales. In the discussion following Theorem~4.3 we argued that the absence of infinite critical droplets coupled with an exponential critical droplet size falloff is inconsistent with a many-state picture.  It might seem that this contradicts the fact that the highly disordered picture, which is known to have infinitely many incongruent states above six dimensions~\cite{JR10}, has only finite critical droplets in all dimensions (though we don't know the distribution of droplet sizes in that model in any dimension greater than one).  In any case, the coupling distribution for the highly disordered model is volume-dependent and therefore violates the assumption behind most of our theorems that the coupling distribution be i.i.d.~over the infinite volume. So no conclusions about the EA model should be drawn from the highly disordered model. (However, we expect that the {\it strongly\/} disordered model should exhibit the same large-lengthscale behavior as the ordinary EA model.)

\bigskip

Turning now to ground state stability, the analysis in this paper suggests two extreme cases. The first corresponds to a ground state having a positive density of bonds which are $\sigma$-critical. If the $\sigma$-criticality is of the first kind, then changing the coupling value of a single bond by~$O(1)$ can completely destroy the structure of the original ground state, leading to a new ground state that is incongruent with the original.  If it is of the second kind, then changing the coupling value of a single bond can lower the flexibility of a positive density of other couplings, making the ground state much less stable to perturbations of arbitrary couplings. It seems appropriate therefore to refer to ground states with a positive density of $\sigma$-critical bonds as ``marginally stable''.\footnote{The term ``marginal stability'' has been used in other contexts, for example, when the Hessian of the interaction potential has a zero eigenvalue; see for example~\cite{FPUZ15}.}

The opposite case is where all critical droplets are finite with $\sigma$-criticality and $\sigma$-subcriticality both absent. In this case a coupling perturbation will lead to at most a local droplet flip, so that any finite number of coupling perturbations leaves the structure of the infinite ground state, in terms of both spin configuration {\it and\/} distribution of flexibilities, essentially intact. As noted above, the latter is guaranteed when a ground state's critical size distribution has an exponential falloff of critical droplet sizes at large lengthscales.  Ground states with this property could justifiably be called ``robustly stable''; the scaling-droplet picture is consistent with this kind of ground state stability.

It is interesting that the two extreme cases --- all ground states being marginally stable vs. all being robustly stable --- correspond to the two opposite cases along the diagonal of Table~1: RSB requires marginally stable ground states, and scaling-droplet likely requires robustly stable ground states. Future work will aim toward further clarifying and refining the picture so far developed.

\acknowledgments We thank Aernout van Enter for helpful suggestions.

\bibliography{refs.bib}

\end{document}